\documentclass[pdflatex,sn-mathphys-ay]{sn-jnl}

\usepackage{graphicx}%
\usepackage{multirow}%
\usepackage{amsmath,amssymb,amsfonts}%
\usepackage{amsthm}%
\usepackage{mathrsfs}%
\usepackage[title]{appendix}%
\usepackage{xcolor}%
\usepackage{textcomp}%
\usepackage{manyfoot}%
\usepackage{booktabs}%
\usepackage{algorithm}%
\usepackage{algorithmicx}%
\usepackage{algpseudocode}%
\usepackage{listings}%
\usepackage{subcaption}

\theoremstyle{thmstyleone}%
\theoremstyle{thmstyletwo}%

\theoremstyle{thmstylethree}%
\begin{document}

\title[Article Title]{Flow mechanisms governing oscillation in a sonic fluidic oscillator}


\author*[1]{\fnm{Chris J.} \sur{Nicholls}}\email{christopher.nicholls@eng.ox.ac.uk}

\author[2]{\fnm{Michael R.} \sur{Fenelon}}\email{michaelfenelon704@gmail.com}

\author[2]{\fnm{Yang} \sur{Zhang}}\email{yzhang215@illinoistech.edu}

\author[2]{\fnm{Louis N.} \sur{Cattafesta III}}\email{lcattafestaiii@illinoistech.edu}

\affil*[1]{\orgdiv{Department of Engineering Science}, \orgname{University of Oxford}, \orgaddress{\street{Parks Road}, \city{Oxford}, \postcode{OX1 3PJ}, \country{United Kingdom}}}

\affil[2]{\orgdiv{Department of Mechanical, Materials, and Aerospace Engineering}, \orgname{Illinois Institute of Technology}, \orgaddress{\street{10 West 32nd Street}, \city{Chicago}, \postcode{60616}, \state{Illinois}, \country{United States}}}

\abstract{
Two factors that influence the oscillation mechanism of a sonic fluidic oscillator are investigated: the geometry of the feedback channel connections (control ports) and the influence of flow restrictions in the oscillator outlets. Phase-averaged planar PIV measurements are performed inside the oscillator, synchronised with unsteady pressure measurements, and analysed using space-only proper orthogonal decomposition (POD). The POD analysis reveals two coupled modes: a Sweeping Mode capturing lateral jet displacement and a Bending Mode capturing jet curvature during switching, the latter being the primary driver of outlet mass flux modulation. Flow separation at the control port entrances is shown to throttle the feedback flow and progressively limit oscillation strength at higher inlet flow rates. Restrictive outlet paths induce a differential back pressure that is shown to cause the jet to separate from its attachment wall and bend towards the splitter tip (`secondary separation'). The secondary separation reduces the differential outlet mass flux and introduces a flow curvature that limits the upstream propagation of the back pressure and thus shields the primary jet attachment. The consequence of these effects is that strong oscillations are sustained down to the smallest outlet apertures investigated. The principal contribution is to demonstrate that the assumed coupling between upstream jet attachment and outlet flow split is broken when the outlet aperture is reduced, with significant implications for the design of fluidic oscillators operating with downstream flow impedances.
}

\maketitle

\section{Introduction}
\label{sec:Introduction}
Fluidic oscillators have attracted significant attention in recent years due to their ability to generate pulsed or sweeping jets without moving parts, making them well-suited to a broad range of flow control applications. In aerodynamics, they have been widely used for active flow control where oscillatory blowing has been shown to be a more efficient means of suppressing separation and increasing lift relative to steady injection \citep{raghu2013fluidic,Seifert1993oscillatory}. Perhaps the most developed use was the Boeing ecoDemonstrator, where fluidic oscillators were employed to enhance rudder effectiveness by delaying flow separation to higher angles of attack \citep{whalen2018flight}. Fluidic oscillators have also found application in thermal management \citep{wu2019large,ghanami2020heat}, aeroacoustics \citep{raman2004cavity}, and combustion control \citep{guyot2008active}. These diverse applications highlight the importance of understanding the internal mechanisms governing oscillations and the influence of downstream conditions, which vary significantly across use cases.

\begin{figure}
\centering
	\begin{subfigure}{0.62\textwidth}
	\includegraphics[width=\textwidth]{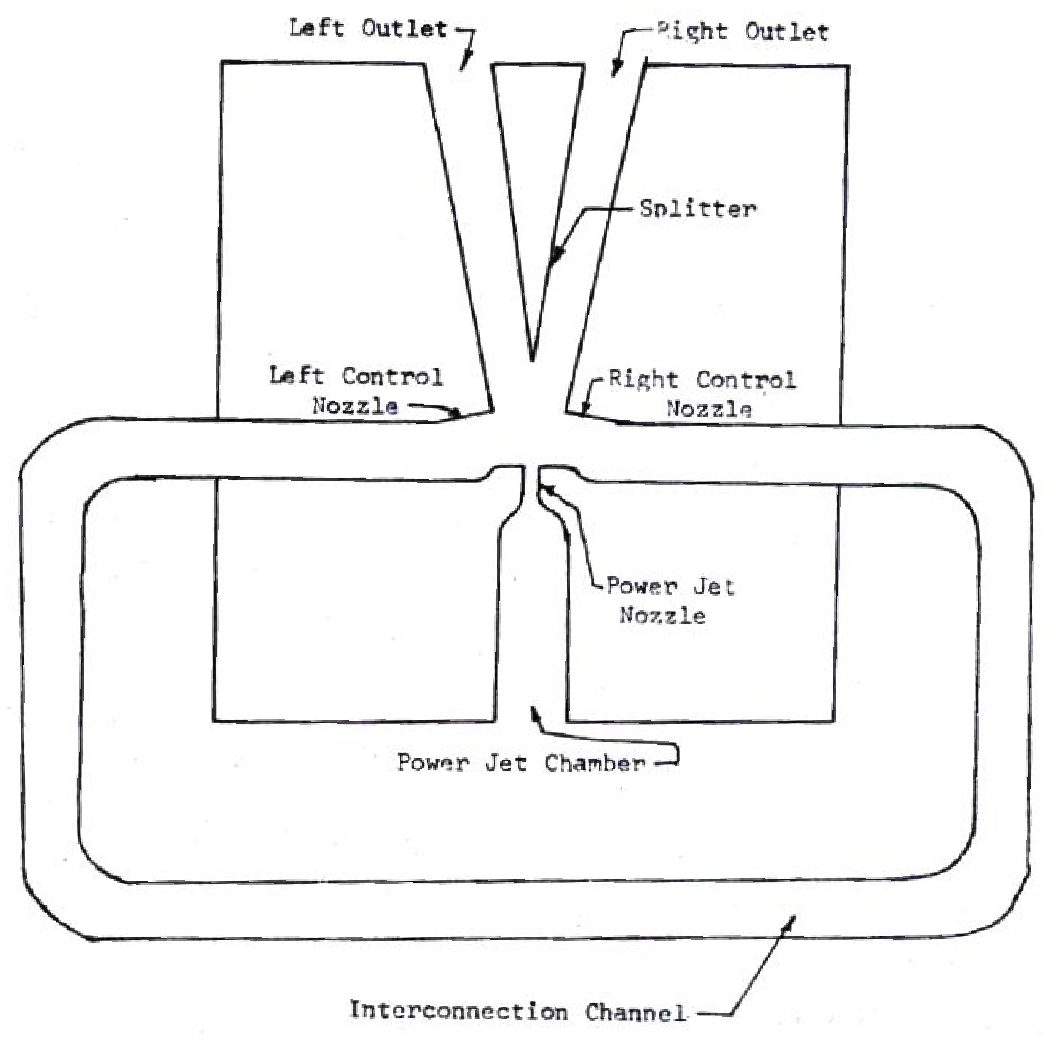}
	\caption{\label{fig:sonic oscillator Spyropoulos}Sonic oscillator, from \citet{spyropoulos1964}}
	\end{subfigure}
	\begin{subfigure}{0.37\textwidth}
	\includegraphics[width=\textwidth]{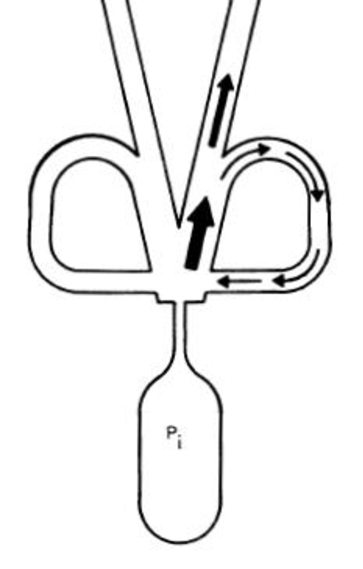}
	\caption{\label{fig:relaxation oscillator}Relaxation oscillator, from \citet{gaylord1969flueric}}
	\end{subfigure}
	\begin{subfigure}{0.67\textwidth}
	\includegraphics[width=\textwidth]{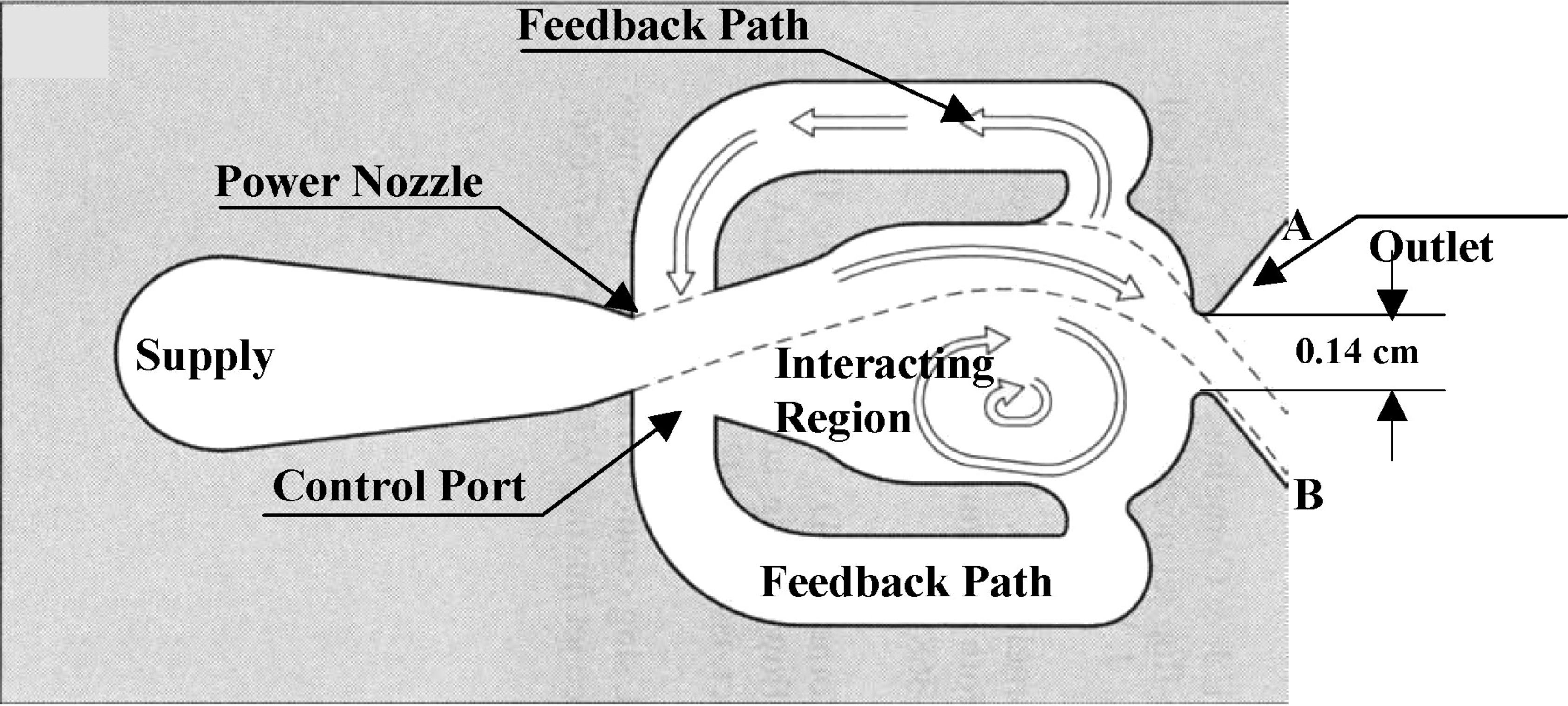}
	\caption{\label{fig:sweeping jet actuator Raman}Sweeping jet actuator, from \citet{raman2004cavity}}
	\end{subfigure}
	\hspace{1em}
	\begin{subfigure}{0.28\textwidth}
	\includegraphics[width=\textwidth]{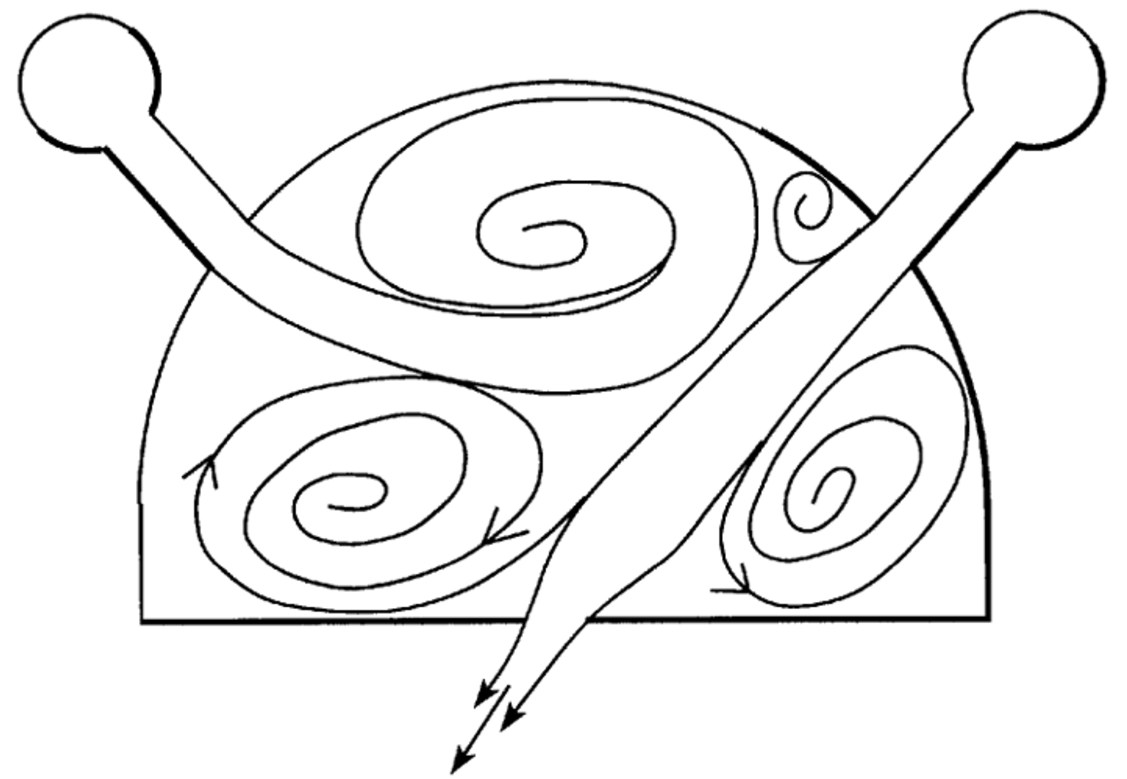}
	\caption{\label{fig:jet interaction oscillator Raghu}Jet interaction oscillator, from \citet{raghu2001feedback}}
	\end{subfigure}
	\caption{\label{fig:fluidic oscillator types}Fluidic oscillator types.}
\end{figure}

Four popular varieties of fluidic oscillator are shown in Fig.\,\ref{fig:fluidic oscillator types}. There is no consensus on terminology for some of these devices, but the names used in the present paper are provided in the captions.
These devices can be categorised as feedback fluidic oscillators (Fig.\,\ref{fig:sonic oscillator Spyropoulos}--\ref{fig:sweeping jet actuator Raman}), because of the use of a channel that introduces an internal feedback that induces the oscillation, or feedback free (Fig.\,\ref{fig:jet interaction oscillator Raghu}). Those shown here using feedback also rely on the Coand\u{a} effect \citep{henri1936device}, the tendency of a flow to attach to an adjacent surface. In these devices, a jet issues from an inlet nozzle into an interaction region where it attaches to one of two side walls that are often set back and/or inclined away from the central axis of the inlet nozzle. In devices that use a splitter (Fig.\,\ref{fig:sonic oscillator Spyropoulos} \& \ref{fig:relaxation oscillator}), the jet is collected into one of two outlet channels; generally the one on the same side to which it is attached upstream. The sonic oscillator (Fig.\,\ref{fig:sonic oscillator Spyropoulos}), first studied by \citet{spyropoulos1964}, is the focus of the present work and is described in detail in Section \ref{sec:Fluidic oscillator behaviour}. The relaxation oscillator (Fig.\,\ref{fig:relaxation oscillator}) and sweeping jet actuator (Fig.\,\ref{fig:sweeping jet actuator Raman}) siphon a portion of the main flow into their feedback paths, while the jet interaction oscillator (Fig\,\ref{fig:jet interaction oscillator Raghu}) requires no Coand\u{a} attachment and operates at frequencies up to 10 kHz \citep{gregory2007characterization}.

Oscillators that employ a splitter to divide the outlet into a pair of pulsed jets (rather than a single, sweeping jet) have the potential to be destabilised by the downstream conditions \citep{nicholls2026two}. A difference in back pressure between the outlet legs has been shown to have a strong influence on the position of the jet inside the device, as opposed to the internal feedback flow(s) as intended \citep{wang2019on}. Indeed, \citet{nicholls2022novel} demonstrated that restrictive and/or long outlet flow paths can induce a bifurcation to a second oscillation mode that is driven by the back pressure, which forms a standing wave along the outlet channels. In a real system, such flow paths connected downstream of an oscillator may be unavoidable. While it has been reported that several groups have observed unexpected and undesirable behaviour when extending oscillator outlets \citep{guyot2008active}, research to explain these effects is scarce.
 
The geometry of the control port connections has not often been considered an important factor in fluidic oscillator design. In almost all designs these connections are oriented perpendicular to the main jet, reflecting the assumption that the control flow acts primarily through its momentum to deflect the jet. However, an alternative mechanism that is based on the aspiration of the recirculation bubble often dominates. In these cases, the direction of the control flow is immaterial and the perpendicular configuration may not be optimal. This distinction is particularly relevant in sonic oscillators, where the bidirectional feedback flow must enter the feedback channel against an aerodynamically unsympathetic turning angle.

This paper addresses two research questions concerning the internal oscillation mechanism of a sonic oscillator, which are developed in detail following a brief review of the relevant background in Section \ref{sec:Fluidic oscillator behaviour}. The experimental methodology is presented in Section \ref{sec:Experimental setup}. Section \ref{sec:Results} presents and analyses the results, with conclusions drawn in Section \ref{sec:Conclusions}.
\section{Background}
\label{sec:Fluidic oscillator behaviour}
\begin{figure}
\centering
 	\begin{subfigure}{0.4\textwidth}
 	\includegraphics[width=\textwidth]{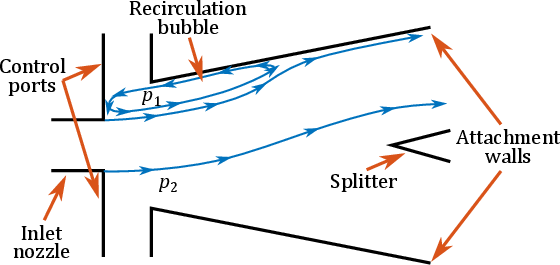}
		\vspace{1.25em}
 	\caption{\label{fig:diverter sketch Coanda explanation}Bistable wall-attachment (fluidic) diverter}
 	\end{subfigure}
 	\begin{subfigure}{0.55\textwidth}
 	\includegraphics[width=\textwidth]{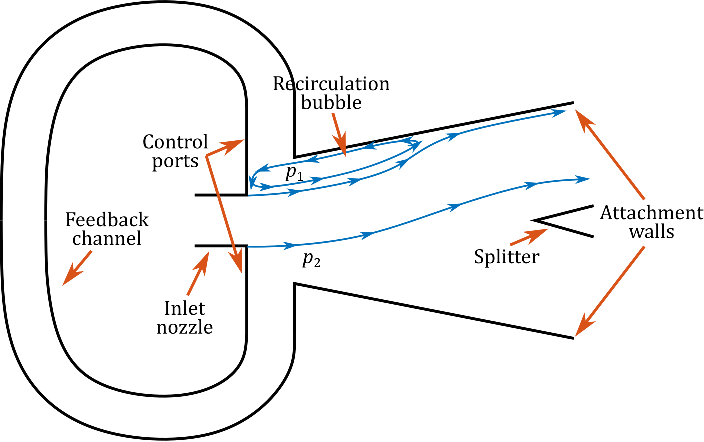}
 	\caption{\label{fig:sonic oscillator sketch}Sonic oscillator \citep{spyropoulos1964}}
 	\end{subfigure}
 	\caption{\label{fig:diverter and sonic oscillator}Fluidic devices.}
\end{figure}

The bistable wall-attachment diverter (Fig.\,\ref{fig:diverter sketch Coanda explanation}) is the core element of several varieties of fluidic oscillator \citep{joyce1983fluidics}. When the supply port is fed from a pressurized source, a jet emerges from the inlet nozzle into the interaction chamber and adheres to one of the sidewalls due to the Coand\u{a} effect \citep{henri1936device}. The jet entrains fluid from the surroundings, reducing the pressure around it due to the confinement provided by the attachment walls. Any minor asymmetry causes the pressure to decrease more on one side than the other, bending the jet towards the more depressed side until it strikes the wall, forming a recirculation bubble. A steady state is reached when the recirculated flow matches the flow entrained into the jet from the bubble \citep{nicholls2023analytical}, with the pressure difference across the jet governed by the balance between the jet's momentum flux per unit depth and its radius of curvature, ${\Delta}p = J/R$.

The jet can be switched by introducing a control flow at the attached side. \citet{muller1964study} identified two switching mechanisms: a \textit{dynamic} mode in which the momentum of the control flow deflects the jet, and a \textit{steady} mode in which the control flow aspirates the recirculation bubble, expanding it until the Coand\u{a} effect is overcome and the jet detaches. The steady mechanism is relevant to sonic oscillators because the feedback channel flow momentum is generally small compared with the main jet; the switching condition reduces to a threshold on the ratio of control mass flow rate to inlet mass flow rate \citep{tippetts1971design, nicholls2026two}.

The sonic oscillator \citep{spyropoulos1964} connects the control ports of a wall-attachment diverter via a single feedback channel (Fig.\,\ref{fig:sonic oscillator sketch}). When the jet attaches to a wall, the resulting sharp pressure drop is transmitted as an expansion wave around the feedback channel, eventually reaching the opposite control port and triggering jet switching. The oscillation period is determined by a combination of the acoustic propagation time in the feedback channel and the jet switching time \citep{spyropoulos1964}. The switching time incorporates the time for the feedback flow to develop to the point where it switches the jet \citep{tippetts1973fluidic}, as well as the time for the jet to traverse the device and reattach to the opposite wall \citep{wassermann2013phase}.
\citet{wassermann2013phase} performed phase-locked three-component MRV measurements inside a sonic oscillator and divided the oscillation cycle into stay, detachment, and attachment phases. They showed that the detachment phase duration scaled with feedback tube length, while the attachment phase duration decreased when the flow rate was increased. The latter provides a possible explanation as to why the oscillation frequency increases with flow rate, which has not yet been addressed in the literature, although more evidence is required to confirm this. Their velocity measurements in the feedback channel are consistent with the aspiration rather than the momentum-switching mechanism being dominant, a point revisited in the present paper.

The influence of downstream conditions on the oscillation mechanism has received comparatively little attention. \citet{guyot2008active} connected tubes to the outlet channels of a fluidic oscillator for a combustion-control application and observed a fourfold reduction in oscillation frequency and a halving of the outlet flow modulation depth. While those authors attributed these effects to the damping of the outlet flows, a back-pressure influence on the oscillation mechanism is a plausible alternative explanation. \citet{nicholls2022novel} demonstrated that restrictive and/or long outlet paths can induce a bifurcation to a distinct oscillation mode driven by a standing wave along the outlet channels, with frequency largely independent of flow rate. \citet{wang2019on} studied a relaxation oscillator numerically and showed that differential back pressure between the outlet channels plays a controlling role in the switching dynamics. At certain phases, the jet separates from its upstream attachment wall and attaches to the opposite wall downstream of the splitter, with switching complete only once sufficient differential back pressure overcomes the downstream attachment. These studies collectively highlight that the assumed direct link between upstream jet attachment and the outlet flow split may break down in the presence of an outlet restriction.

In almost all fluidic oscillator designs, the feedback or control channels connect to the interaction region perpendicular to the main jet, reflecting the assumption that control flow momentum is the primary switching agent. However, where the steady (aspiration) mechanism dominates, the direction of the control flow is immaterial and this geometry may not be optimal. In a sonic oscillator the feedback flow is bidirectional, meaning the flow path from the interaction region into the feedback channel is aerodynamically unsympathetic in the conventional perpendicular configuration. The consequences of this feature have not previously been studied.

The present paper addresses two open questions. First, what is the effect of the perpendicular control port geometry on the feedback channel flow and the oscillation mechanism? Second, how does a differential back pressure induced by restrictive outlet apertures influence the oscillation?
The present paper addresses both questions through phase-averaged planar PIV measurements of the internal flow in a sonic oscillator synchronized with unsteady pressure measurements.
\section{Experimental setup}
\label{sec:Experimental setup}
\subsection{Device geometry}
\label{sec:Device geometries}
\begin{figure}[h]
    \centering
    %
    \begin{subfigure}{0.48\textwidth}
    \includegraphics[width=\textwidth]{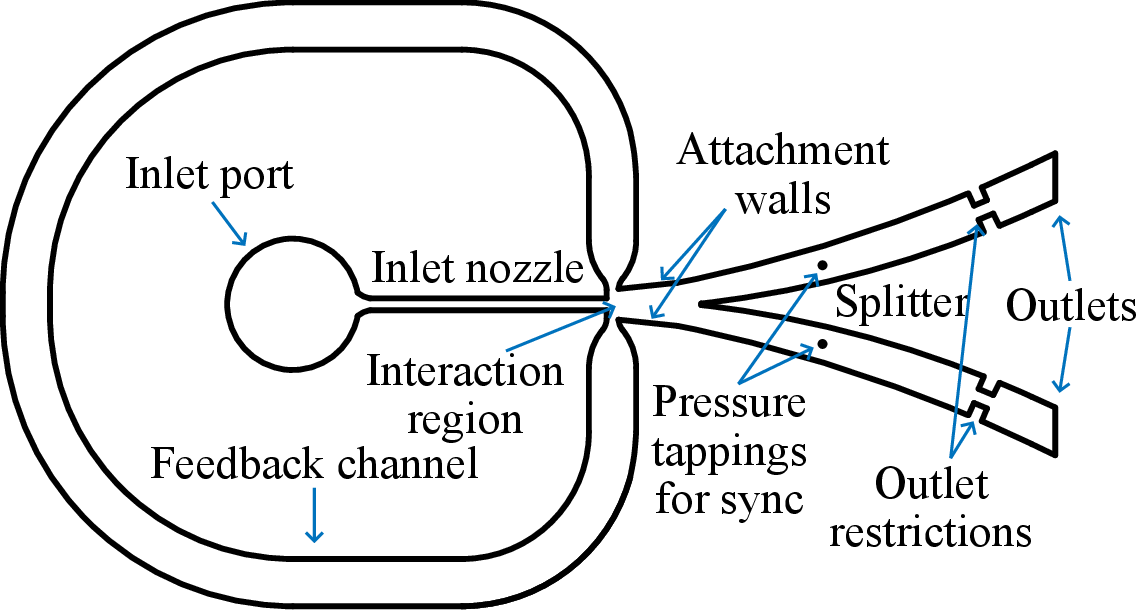}
    \caption{\label{fig:device geometry labelled}Labelled diagram}
    \end{subfigure}
    \hfill
    \begin{subfigure}{0.48\textwidth}
    \includegraphics[width=\textwidth]{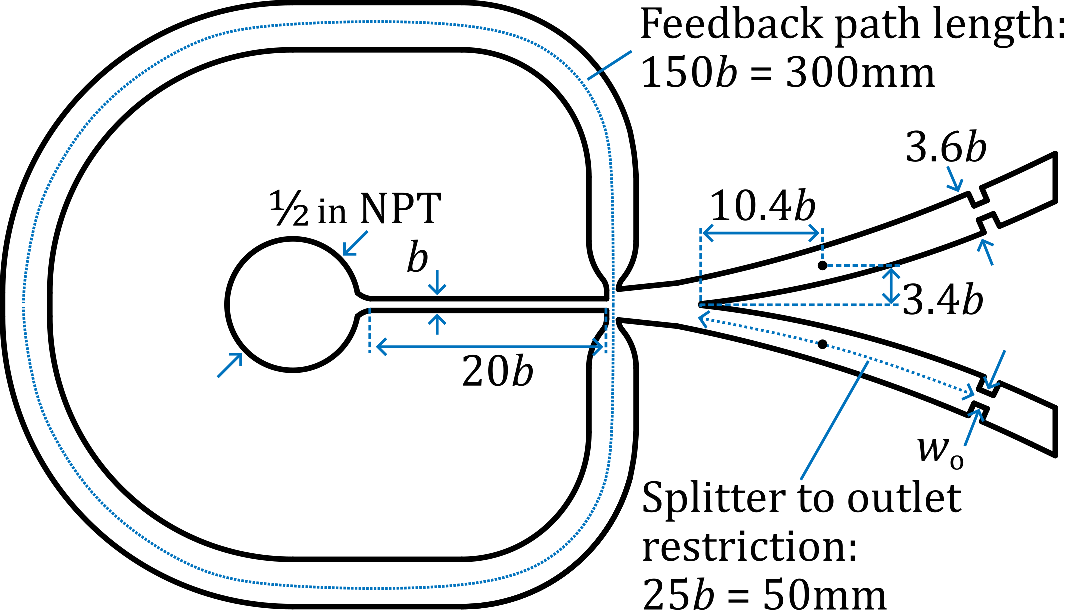}
    \caption{\label{fig:device geometry dimensions}Flow path dimensions}
    \end{subfigure}
    \vspace{0.5em}
    %
    \begin{subfigure}{0.38\textwidth}
    \includegraphics[width=\textwidth]{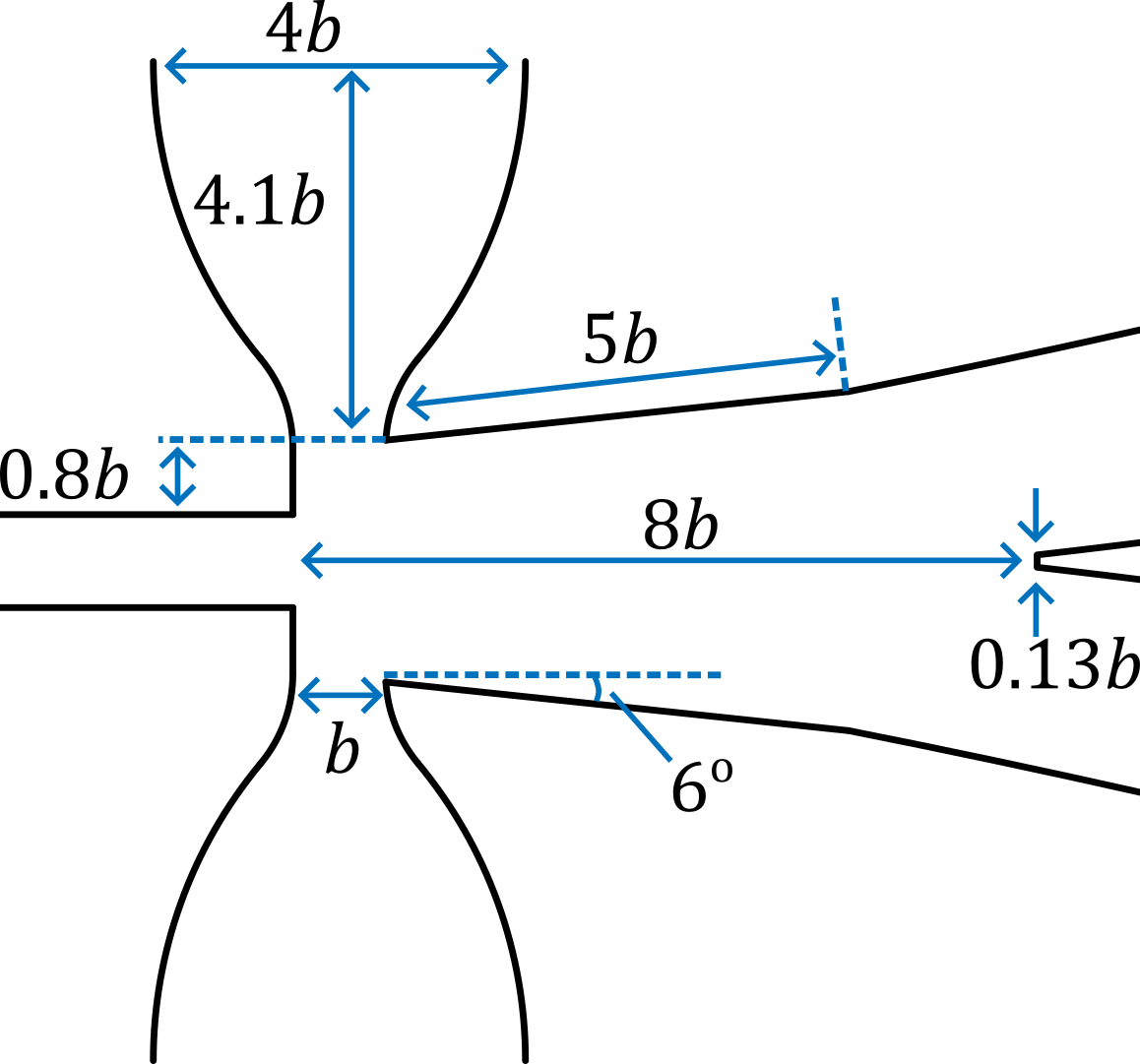}
    \caption{\label{fig:device geometry dimensions zoom}Flow path dimensions (detail)}
    \end{subfigure}
    \hfill
    \begin{subfigure}{0.48\textwidth}
    \includegraphics[width=\textwidth]{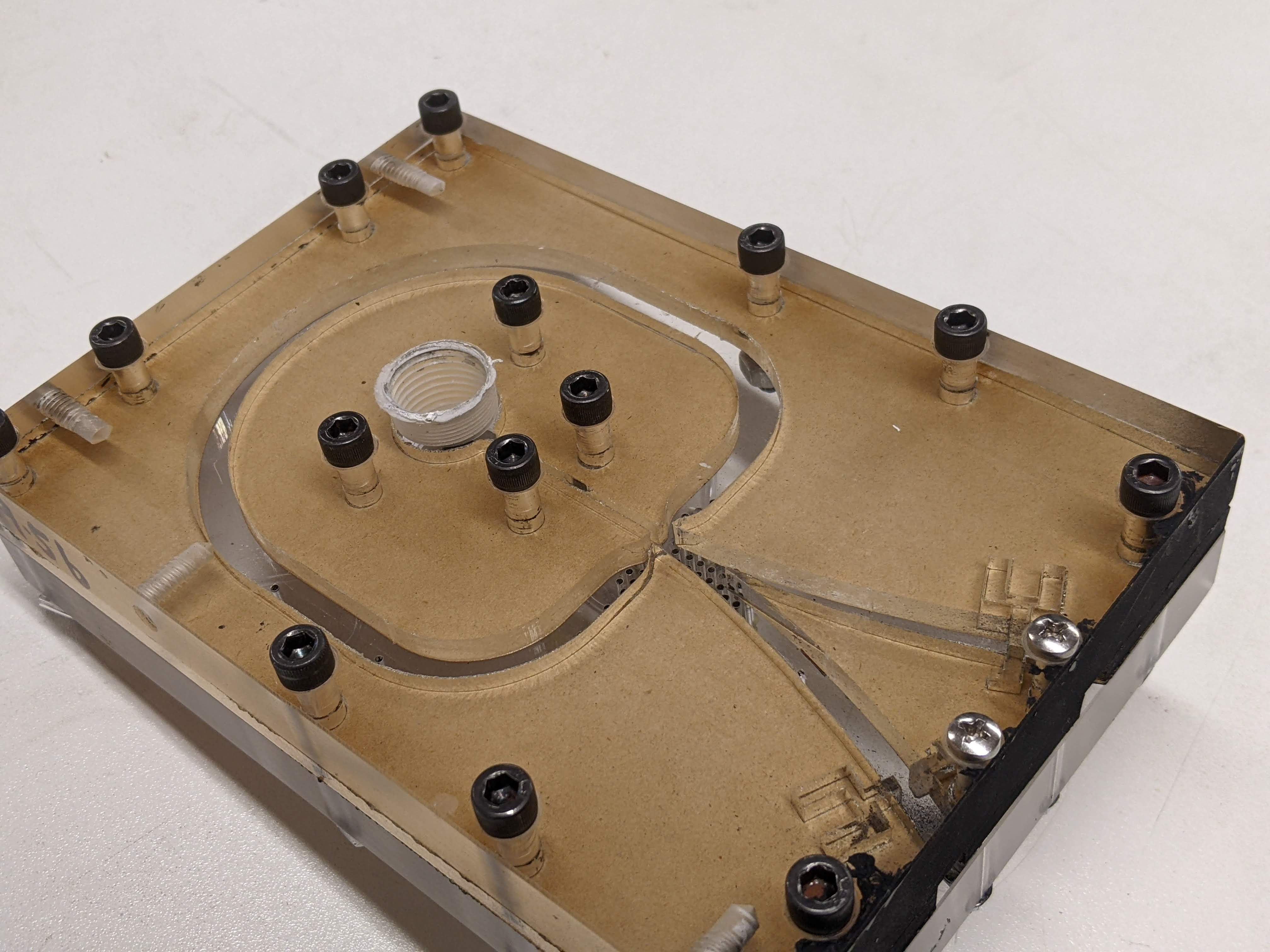}
    \caption{\label{fig:device photograph}Photograph of assembled device}
    \end{subfigure}
    \caption{Oscillator geometry: (a) labelled diagram; (b) \& (c) flow path dimensions; (d) photograph of assembled device. Inlet nozzle width $b=2$\,mm.}
    \label{fig:device geometries}
\end{figure}

Fig.\,\ref{fig:device geometries} shows the geometry and key dimensions of the oscillator, together with a photograph of the assembled device. The inlet port is supplied out of plane in both directions by a symmetric supply to avoid introducing an out-of-plane velocity. The width of the inlet nozzle is $b=2$\,mm. The flow path was laser cut from $6.4$\,mm acrylic, which was adhered to a bottom plate ($17.5$\,mm acrylic) using a low surface energy transfer tape (3M 8132LE). Since laser cutters are only capable of through-cuts, the regions enclosed by the feedback path and splitter are left as `unsupported islands.' Small geometric asymmetries in a fluidic device introduce a bias to its operation, so thin bridges were introduced to align these islands with the surrounding acrylic; these were removed once the flow path was bonded to the bottom plate. A top plate ($17.5$\,mm acrylic) was then bolted to the now-combined flow path and bottom plate, and sealed with gasket paper laser-cut to the negative of the flow path. A variable restriction was introduced to the outlet path in the form of a sharp contraction with aperture width $w_\text{o}$ (Fig.\,\ref{fig:device geometries}(c)); restrictions of varying sizes were 3D printed and installed in a keyed section cut from the surrounding acrylic. The outlet channel width immediately upstream of the outlet restrictions is $3.6\,b = 7.2$\,mm, producing contraction ratios of $3.6b/w_\text{o}$. A photograph of the assembled device is shown in Fig.\,\ref{fig:device geometries}(d).
\subsection{Experimental arrangement}
\begin{figure}[t]
    \centering    
    \includegraphics[width=0.65\textwidth]{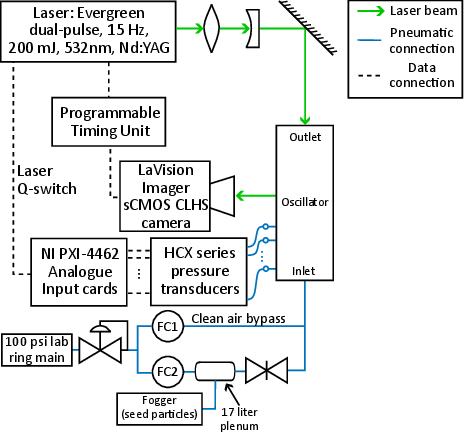}
    \caption{Setup for PIV experiments.}
    \label{fig:PIV setup}
\end{figure}

The setup used for the PIV experiments is shown in Fig.\,\ref{fig:PIV setup}, where FC1 (Omega FMA-2612A, 0--10.6\,g/s) and FC2 (Omega FMA-2609A, 0--1.06\,g/s) control the clean air bypass and seeded flows, respectively. 
Water-based seed particles produced by a fogger (ADJ VF400) were introduced into a $17$-litre plenum. The seeded flow was mixed with a clean air bypass, and the combined flow split symmetrically via a tee-section to supply the two device inlets; care was taken to match the hose lengths on each side. Pressure tappings in the outlet channels (Fig.\,\ref{fig:device geometries}) were connected to $0$--$50$\,mbar pressure transducers (HCXM050D6V), whose analogue outputs were sampled at $50$\,kHz by a National Instruments PXI chassis via an NI PXI-4462 analogue input card. The laser Q-switch signal was also sampled, allowing for synchronisation between the image and pressure data. The laser (two Evergreen dual-pulse Nd:YAG 532\,nm, 200\,mJ, 15\,Hz) and a LaVision Imager sCMOS CLHS camera ($2560 \times 2160$ pixels) were controlled by a programmable timing unit.
\subsection{Phase-averaging}
While the flow field of a fluidic oscillator is periodic and statistically stationary, the oscillation is subject to significant phase jitter -- that is,
the oscillation frequency varies from cycle to cycle due to the inherently random turbulent fluctuations that drive the entrainment mechanism. The phase of each PIV snapshot must therefore be assigned individually rather than inferred from a fixed sampling rate.

In the present work, the phase binning strategy was based on time-resolved pressure measurements. The phase of the pressure signal oscillation at the time of the snapshot was obtained by cross-correlating the period of the corresponding pressure signal with that of a reference pressure signal. The reference pressure signal was computed by finding the positive-slope zero-crossing of each pressure cycle and ensemble-averaging over all cycles, producing a representative reference waveform. Snapshots were sorted into one of 30 bins, giving a 12$^\text{o}$ bin width. The cycle-to-cycle standard deviation of the oscillation period was 8.1\,\% of the mean period. For the primary dataset (0.70\,g/s, 2000 total snapshots), this yields approximately 67 snapshots per phase bin.
\subsection{Scope of results}
\label{sec:Scope of results}
Table \ref{tab:Summary of experiments} summarises the experiments, which comprise two series: varying inlet mass flow rate at a fixed outlet aperture
of $w_\text{o} = 2.5$\,mm, and varying outlet aperture at a fixed flow rate of $\dot{m} = 0.70$\,g/s. At $w_\text{o} = 5.0$\,mm the jet remained stably attached to one wall with only a weak oscillation superposed; two separate measurements were therefore made with the jet
attached to the upper and lower walls respectively, as indicated in the table. The interaction region and the feedback channel were measured in separate experiments using different laser inter-pulse times, owing to the order-of-magnitude difference in flow speed between the two regions. Phase alignment between the two measurement domains relies on the pressure-based synchronisation described earlier.
\begin{table}[h]
\caption{Summary of experiments.}
\centering
\begin{tabular}{cccc}
\toprule
Outlet aperture, $w_\text{o}$ [mm] & \# images & $\dot{m}$ [g/s] & $f_\text{osc}$ [Hz] \\
\midrule
$2.5$ & 1000 & 0.49 & 51  \\
$2.5$ & 2000 & 0.70 & 76  \\
$2.5$ & 1000 & 0.91 & 99  \\
$2.5$ & 500  & 1.8  & 194 \\
\midrule
$5.0$, upper & 1000 & 0.70 & ${\sim}96^\dagger$  \\
$5.0$, lower & 1000 & 0.70 & ${\sim}84^\dagger$  \\
$3.5$        & 1000 & 0.70 & 78 \\
$2.0$        & 1000 & 0.70 & 84 \\
$1.4$        & 1000 & 0.70 & 88 \\
$0.9$        & 500  & 0.70 & 91 \\
$0.5$        & 1000 & 0.70 & 91 \\
\bottomrule
\multicolumn{4}{l}{$^\dagger$ Approximate; broad spectral peak due to stable jet attachment with weak superposed oscillation at this aperture.}
\end{tabular}
\label{tab:Summary of experiments}
\end{table}
\section{Results}
\label{sec:Results}
\subsection{Description of flow field}
\label{sec:Description of flow field}
\begin{figure}
    \centering    
    \includegraphics[width=0.75\textwidth]{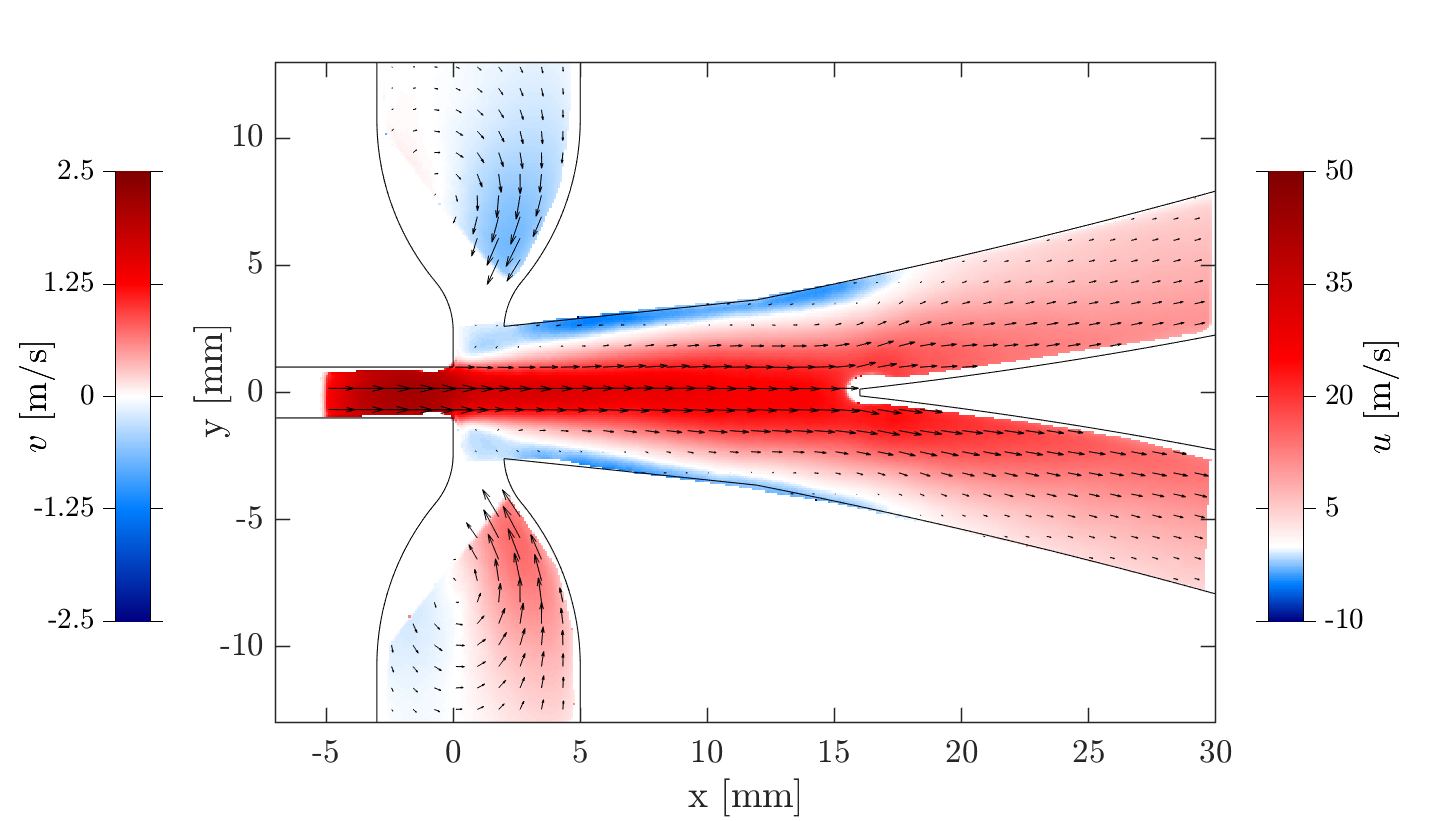}
    \caption{Mean flow (ensemble average of phase bins) at $0.70$\,g/s, $w_\text{o}=2.5$\,mm. The interaction region shows u-velocity with the right-hand colour bar; the feedback channel regions show v-velocity with the left-hand colour bar. Quiver scalings differ between the two regions. The outlet restrictions are further downstream than the visible region.}
    \label{fig:mean flow}
\end{figure}

The mean flow for $w_\text{o}=2.5$\,mm at $0.70$\,g/s, computed by averaging 2000 snapshots, is shown in Fig.\,\ref{fig:mean flow}. The oscillation frequency at this condition was $77$\,Hz. In Fig.\,\ref{fig:mean flow} and all subsequent flow field figures, the interaction region and feedback channel are shown with separate colour bars and quiver scalings (see caption); the upper and lower feedback channel entrances are referred to as the upper and lower feedback channels, respectively. The mean flow shows a jet that spreads and splits between the two outlet channels, with a slight bias towards the lower outlet channel attributed to a minor geometric imperfection, consistent across repeated experiments. The feedback channel entrances show a circulating flow in the mean, with an in-wash on the downstream side and an out-wash on the upstream side, discussed in detail in Section\,\ref{sec:Circulation in control ports}.

\begin{figure}[H]
    \centering
    %
    \begin{subfigure}{0.49\textwidth}
    \includegraphics[width=\textwidth]{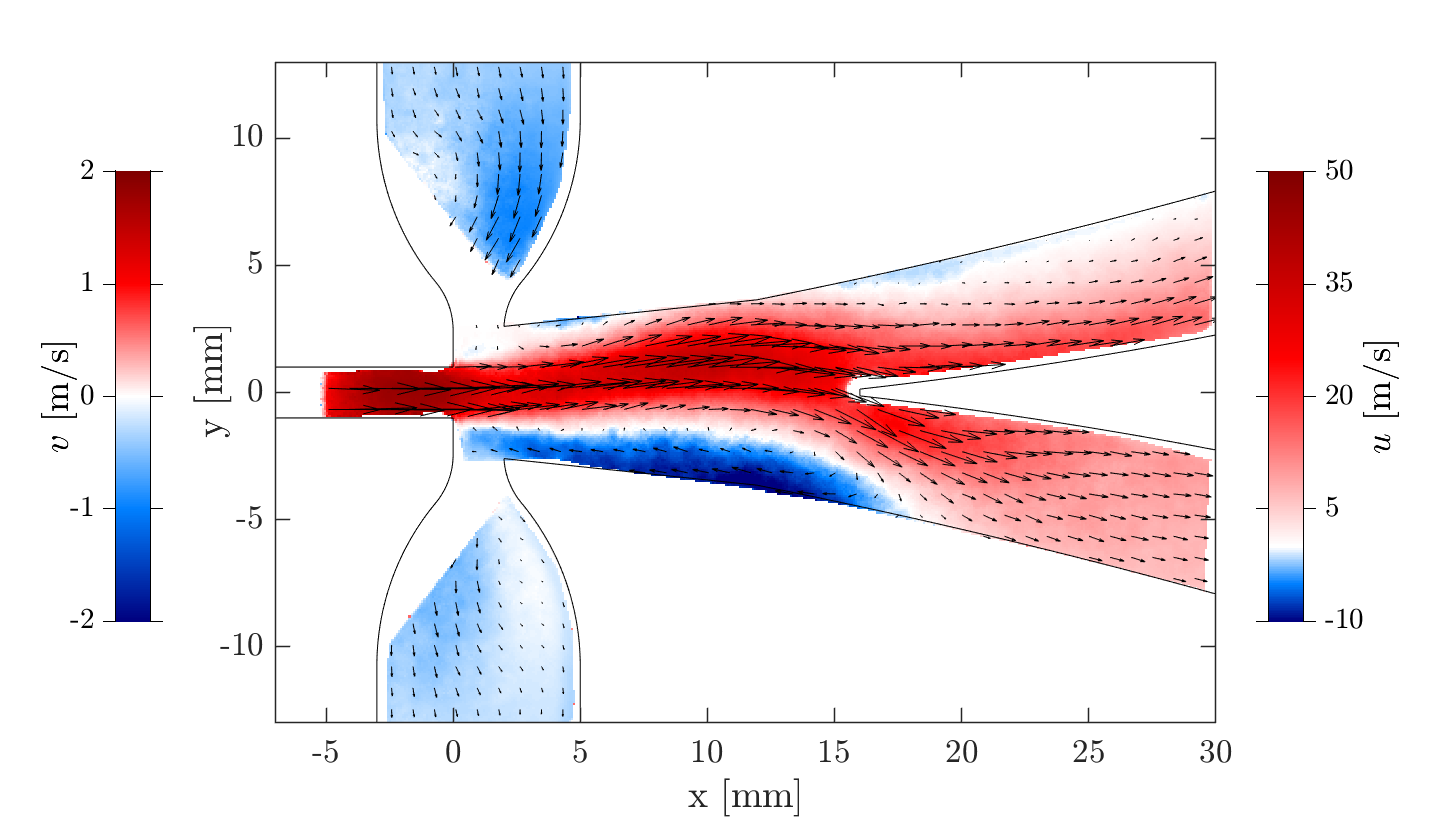}
    \caption{${\phi}=0^\text{o}$}
    \end{subfigure}
    \hfill
    \begin{subfigure}{0.49\textwidth}
    \includegraphics[width=\textwidth]{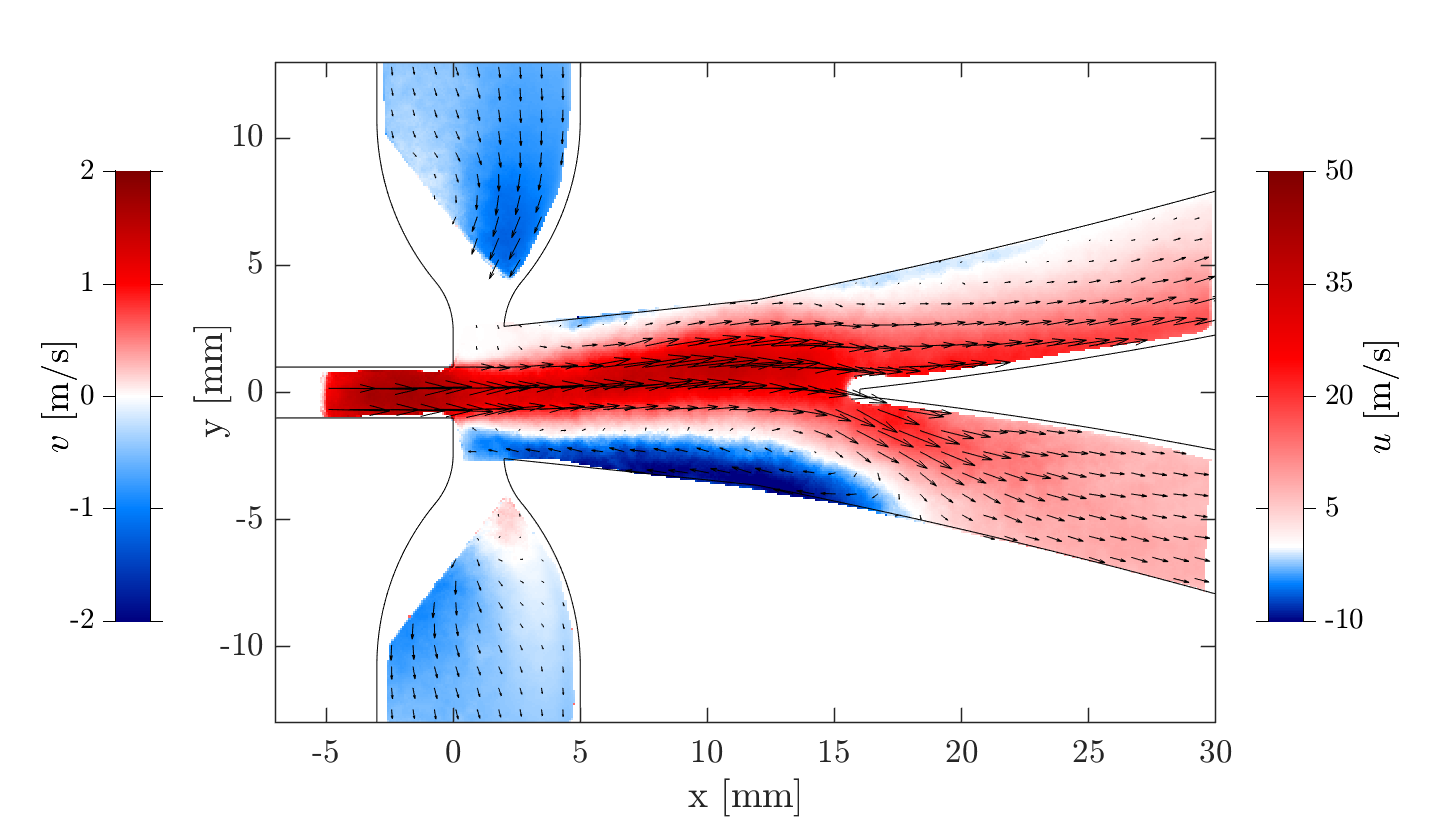}
    \caption{${\phi}=36^\text{o}$}
    \end{subfigure}
    \vspace{0.5em}
    %
    \begin{subfigure}{0.49\textwidth}
    \includegraphics[width=\textwidth]{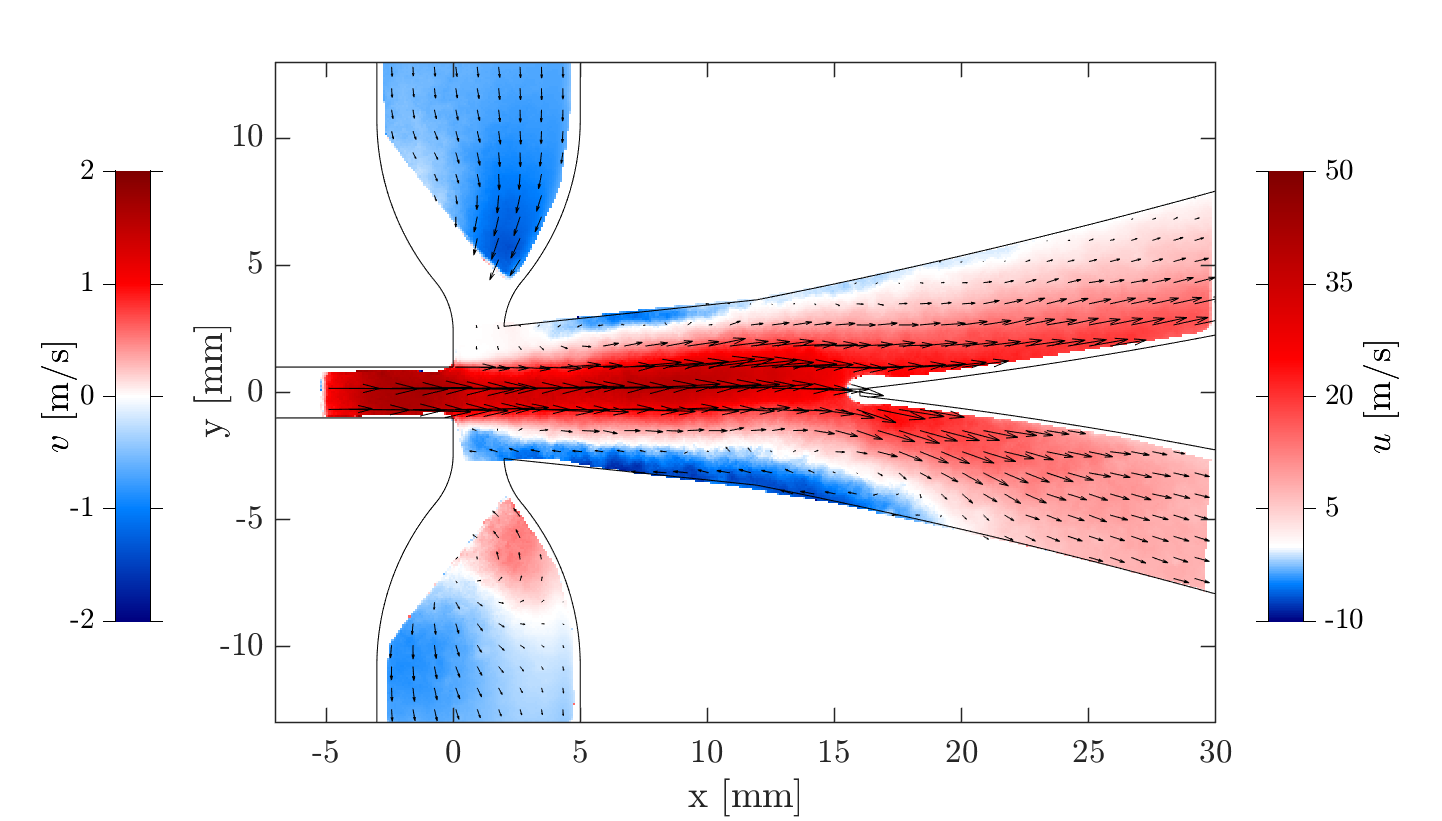}
    \caption{${\phi}=72^\text{o}$}
    \end{subfigure}
    \hfill
    \begin{subfigure}{0.49\textwidth}
    \includegraphics[width=\textwidth]{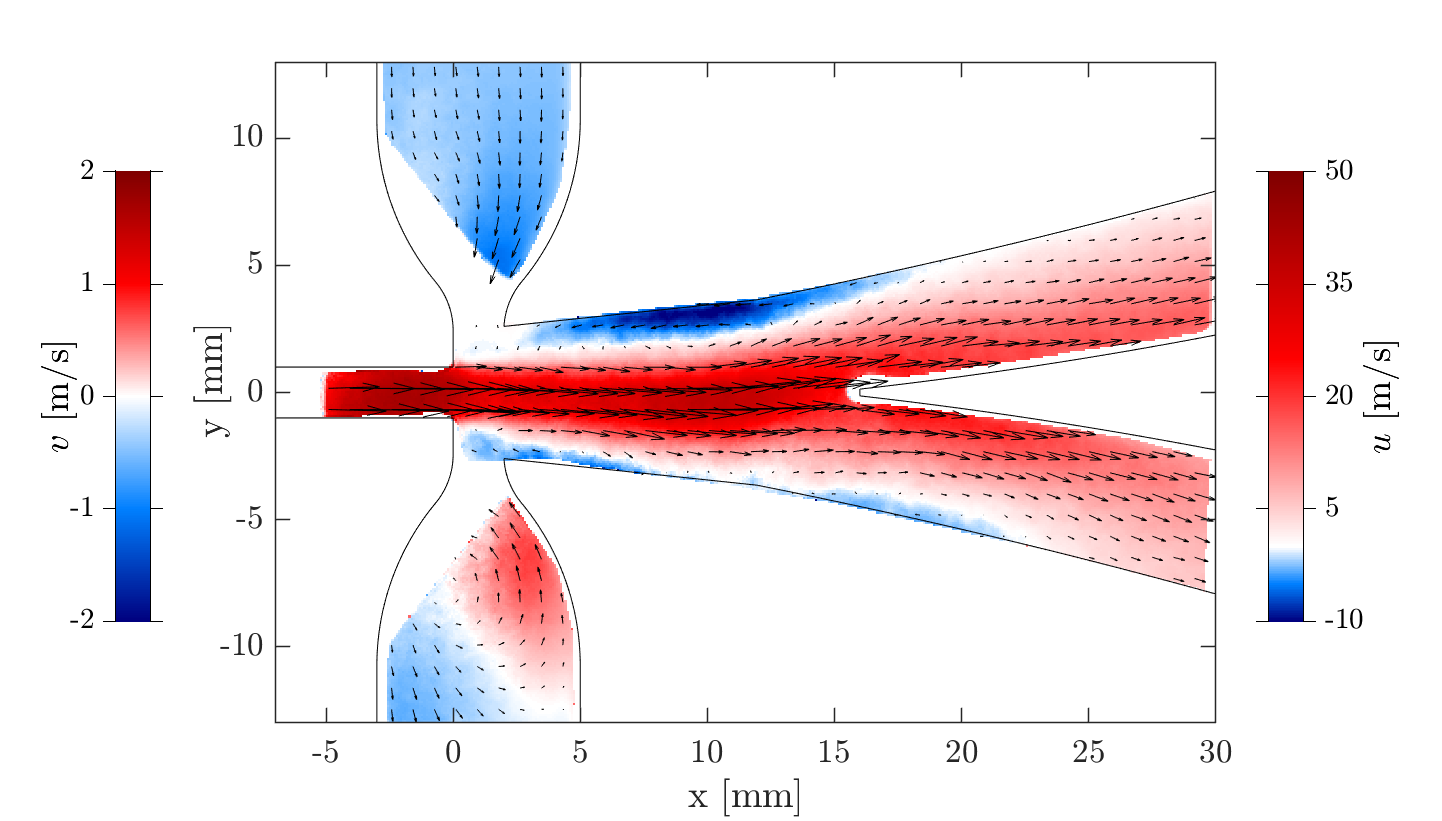}
    \caption{${\phi}=108^\text{o}$}
    \end{subfigure}
    \vspace{0.5em}
    %
    \begin{subfigure}{0.49\textwidth}
    \includegraphics[width=\textwidth]{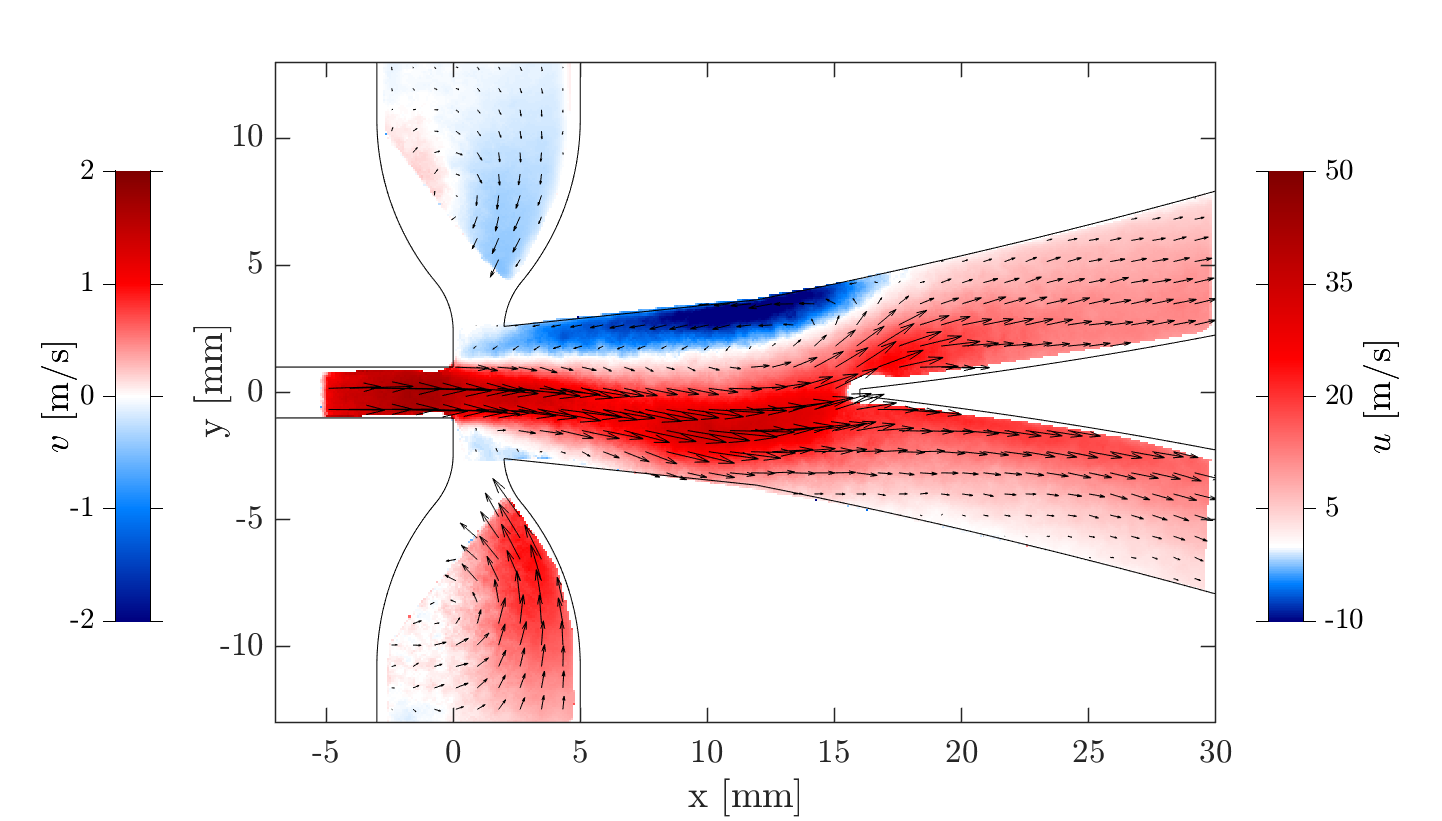}
    \caption{${\phi}=144^\text{o}$}
    \end{subfigure}
    \hfill
    \begin{subfigure}{0.49\textwidth}
    \includegraphics[width=\textwidth]{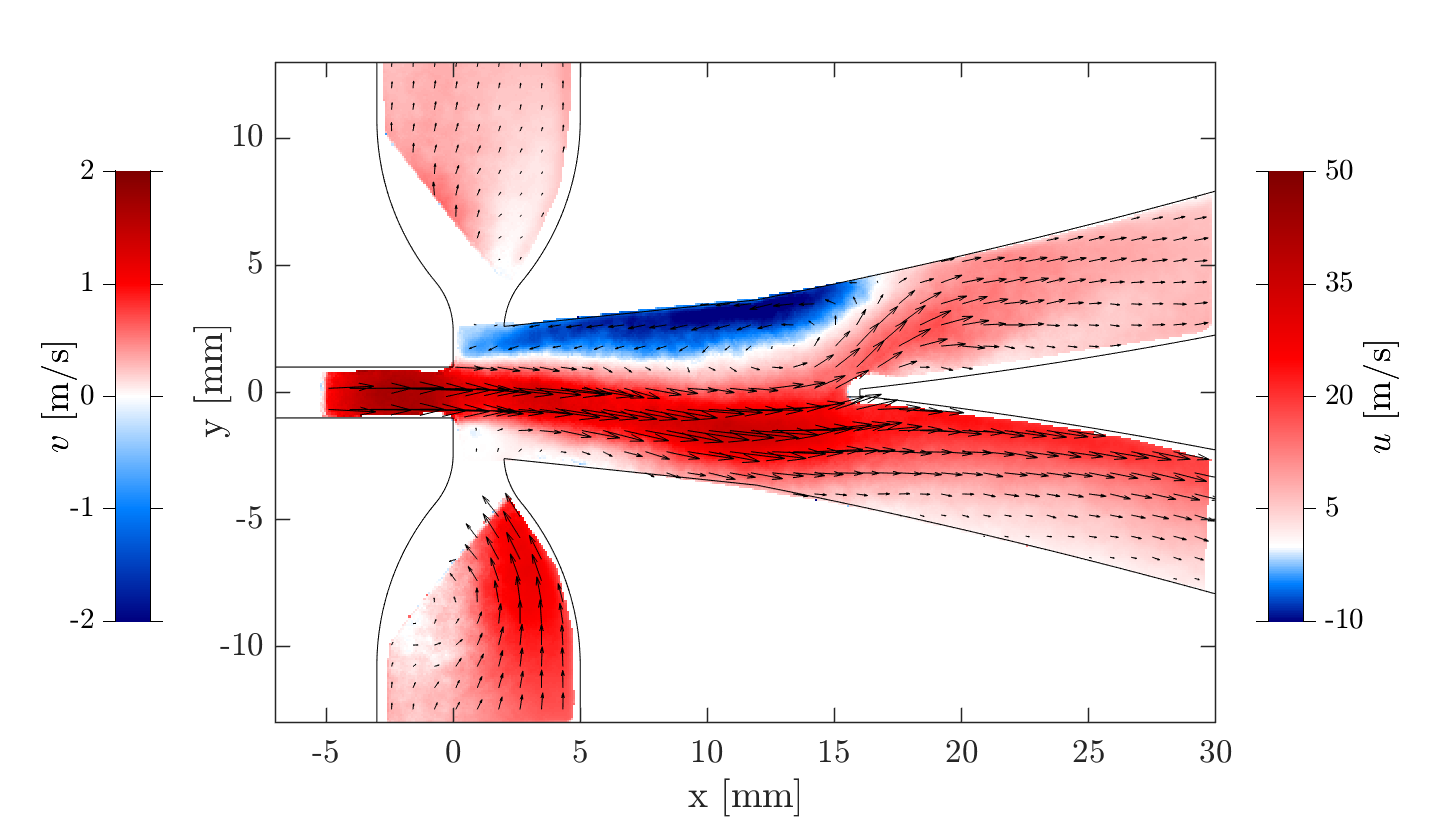}
    \caption{${\phi}=180^\text{o}$}
    \end{subfigure}
    \vspace{0.5em}
    %
    \begin{subfigure}{0.49\textwidth}
    \includegraphics[width=\textwidth]{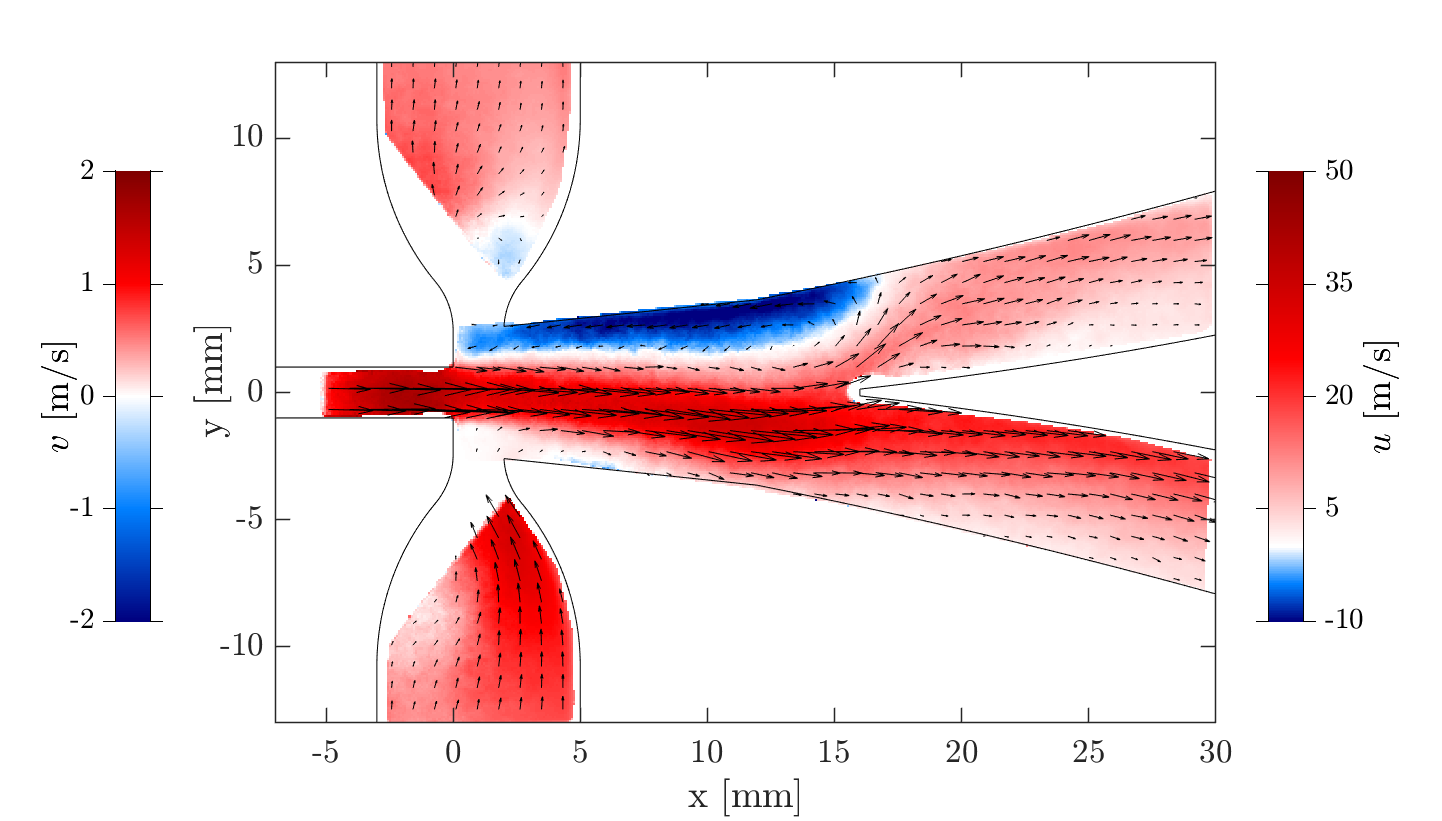}
    \caption{${\phi}=216^\text{o}$}
    \end{subfigure}
    \hfill
    \begin{subfigure}{0.49\textwidth}
    \includegraphics[width=\textwidth]{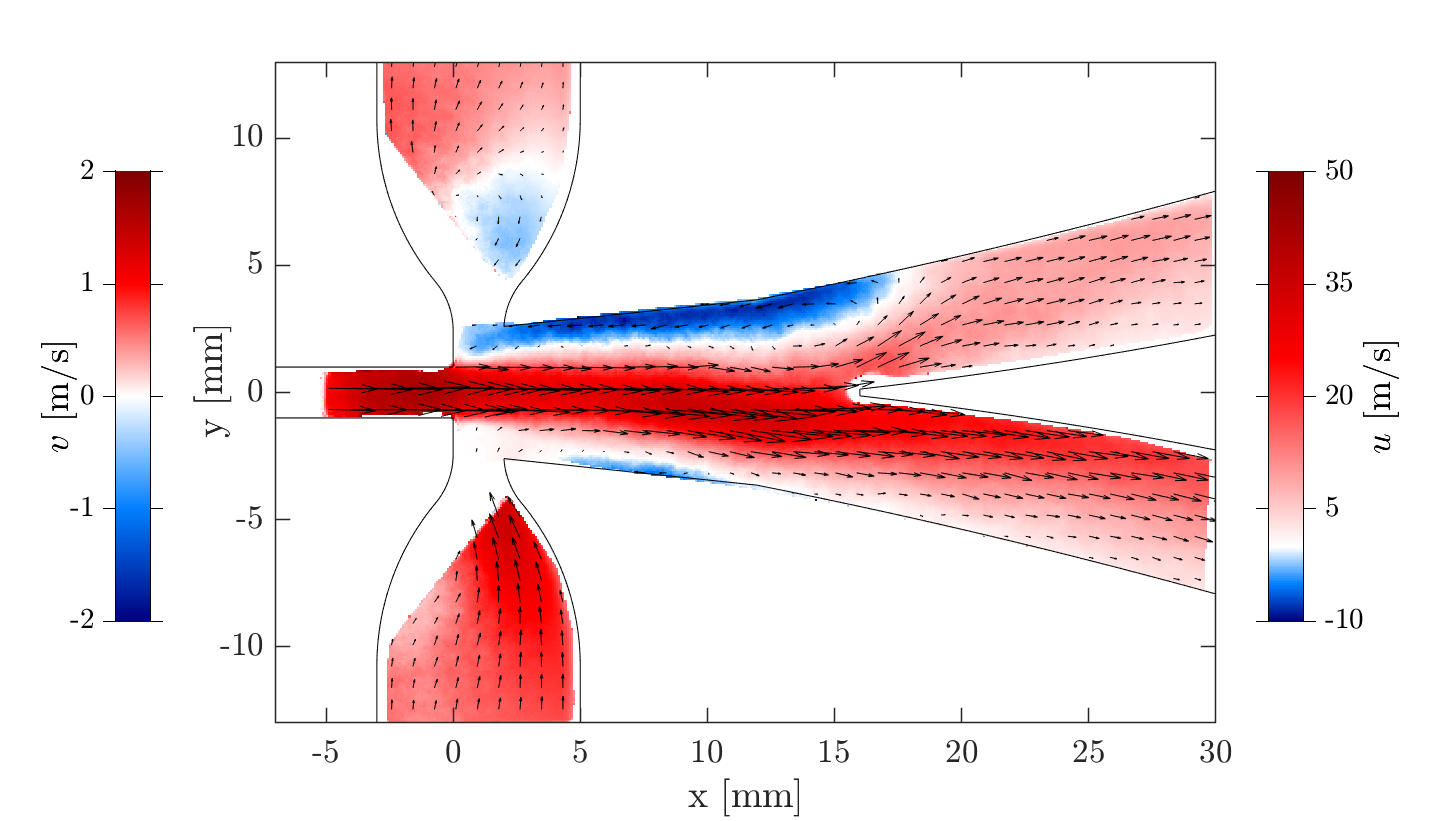}
    \caption{${\phi}=252^\text{o}$}
    \end{subfigure}
    \vspace{0.5em}
    %
    \begin{subfigure}{0.49\textwidth}
    \includegraphics[width=\textwidth]{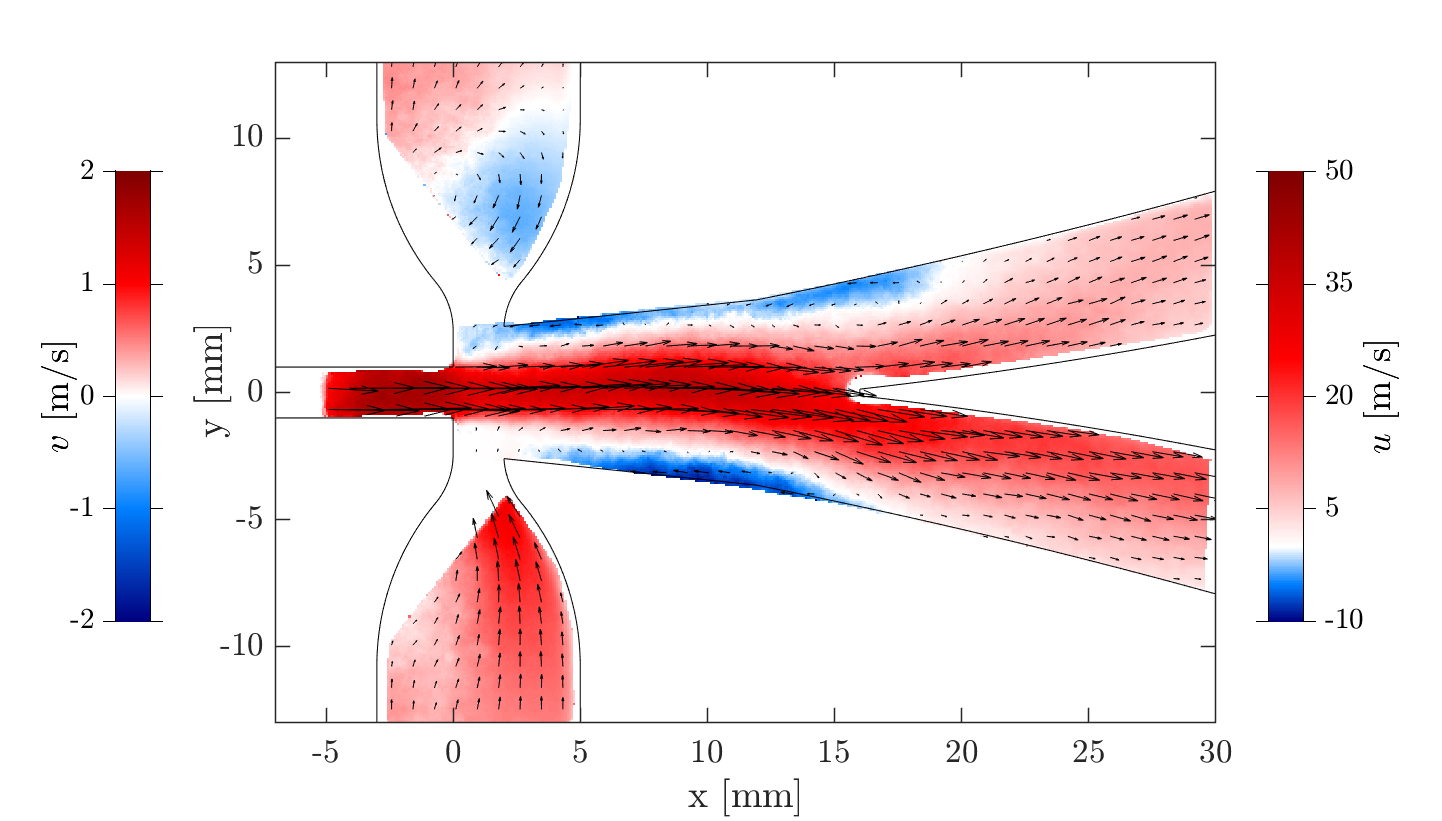}
    \caption{${\phi}=288^\text{o}$}
    \end{subfigure}
    \hfill
    \begin{subfigure}{0.49\textwidth}
    \includegraphics[width=\textwidth]{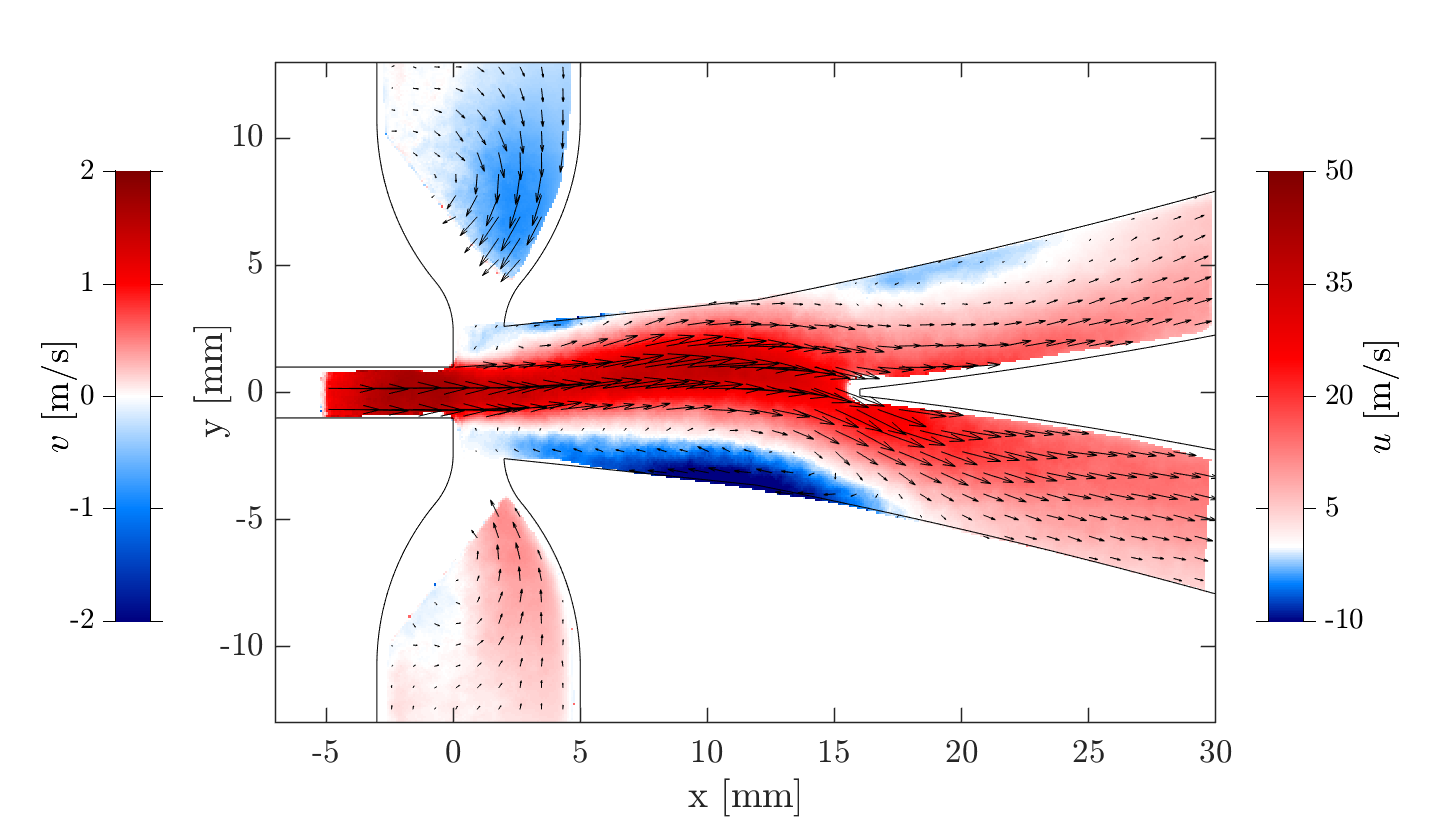}
    \caption{${\phi}=324^\text{o}$}
    \end{subfigure}
    \caption{Phase-averaged 2D PIV inside device A-$2.5$ at $0.70$\,g/s.}
    \label{fig:exp34 to 41 2D PIV inside oscillator}
\end{figure}

Fig.\,\ref{fig:exp34 to 41 2D PIV inside oscillator} shows the phase-averaged data at the same condition, with every third phase bin shown for brevity, illustrating the typical sweeping jet motion. In this geometry, the jet tends to attach to the splitter so that the flow entering each outlet channel is a steady jet with a superposed, sinusoidal pulsing. In Fig.\,\ref{fig:exp34 to 41 2D PIV inside oscillator}(a), the jet is attached to the upper side of the device, close to its extremum position. The process of detachment has begun in (b), where the jet has shifted very slightly away from the wall between the nozzle orifice and the reattachment point. This is associated with an increase in the size of the recirculation bubble between the jet and the upper attachment wall immediately downstream of the nozzle exit. Subfigures (a) and (b) also show a down-wash in both upper and lower feedback channels (flowing from the lower side around to the upper side), which supplies the fluid that is expanding the recirculation bubble. 
In addition to the upstream, primary separation bubble, a secondary flow separation from the upper wall that develops downstream at around $x = 15$\,mm can be seen in (a)--(c) and is clearest in (j). This is a result of the jet bending downwards from the upper wall to attach to the splitter. It is encouraged by the sharp change in the upper wall curvature at $x = 12$\,mm, immediately upstream of the separation. As the jet peels off the wall between (a) and (c), the secondary separation bubble elongates and collapses. The secondary separation bubble is discussed in detail in Section \ref{sec:Effect of varying outlet restriction}.

By (d), the primary separation bubble has expanded sufficiently that the jet has detached from the upper wall and is traversing the device towards the lower wall, with the reattachment point now far downstream in the outlet channel. The jet has almost reached the lower wall and the primary and secondary separation bubbles on the lower side have formed. As the lower-wall attachment strengthens from (d) to (e), jet entrainment evacuates the lower primary separation bubble, drawing an increasing flow from the lower feedback channel. Subfigures (f) to (j) then mirror (a) to (e) with the jet starting attached to the lower wall.

The phase-averaged data also reveal a delay in the feedback channel response between the initiation of the up-wash in the lower feedback channel and its appearance in the upper feedback channel in (d)--(f). This delay supports the notion that the feedback channel flow is initiated on the side to which the jet attaches rather than on the unattached side. In other words, while attached to the lower wall as in (d)--(f), the flow is sucked around the feedback channel towards the lower side rather than blown from the other end.
As described by \citet{spyropoulos1964}, the evacuation of the lower separation bubble reduces the pressure there sharply, producing an expansion wave that travels around the feedback channel to the upper side. Flow is only drawn into the upper entrance of the feedback channel from the interaction region once the expansion wave arrives there. The delay, most clearly visible with the mean flow removed from the feedback channel 
$v$-velocity contour, is approximately two phase bins in magnitude, or 
$0.87 \pm 0.21$\,ms. For a $300$\,mm feedback channel at a sonic speed of 
$340$\,m/s, the expected acoustic propagation delay is $0.88$\,ms --- within the 
measurement uncertainty and consistent with the acoustic wave propagation mechanism 
proposed by \citet{spyropoulos1964}.
\subsection{POD analysis}
\label{sec:POD analysis}
\begin{figure}
    \centering
    \includegraphics[width=0.75\textwidth]{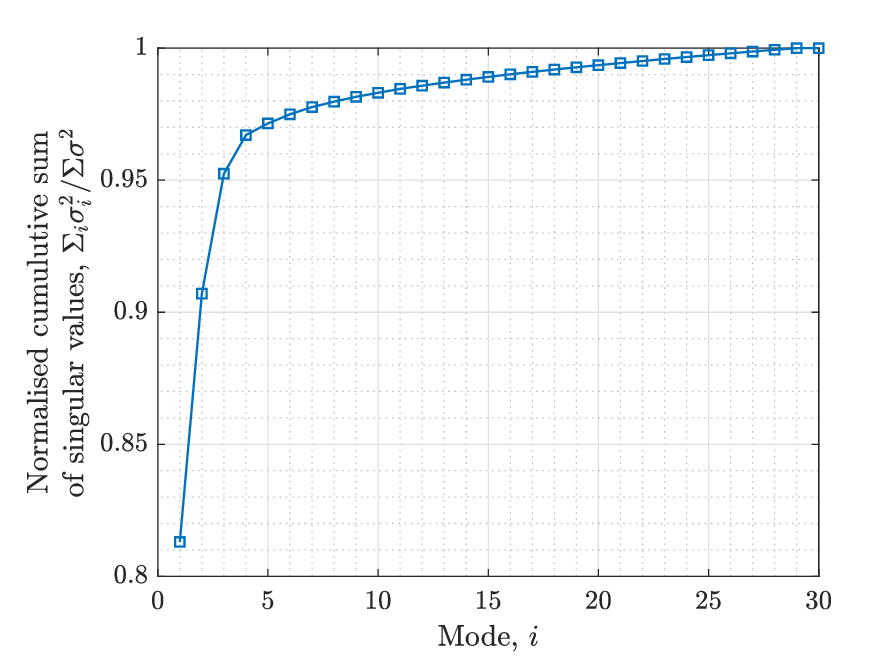}
    \caption{Cumulative sum of squared singular values, normalised by variance.}
    \label{fig:POD singular cumulative}
\end{figure}

\begin{figure}
\centering
    \begin{subfigure}{0.49\textwidth}
    \includegraphics[width=\textwidth]{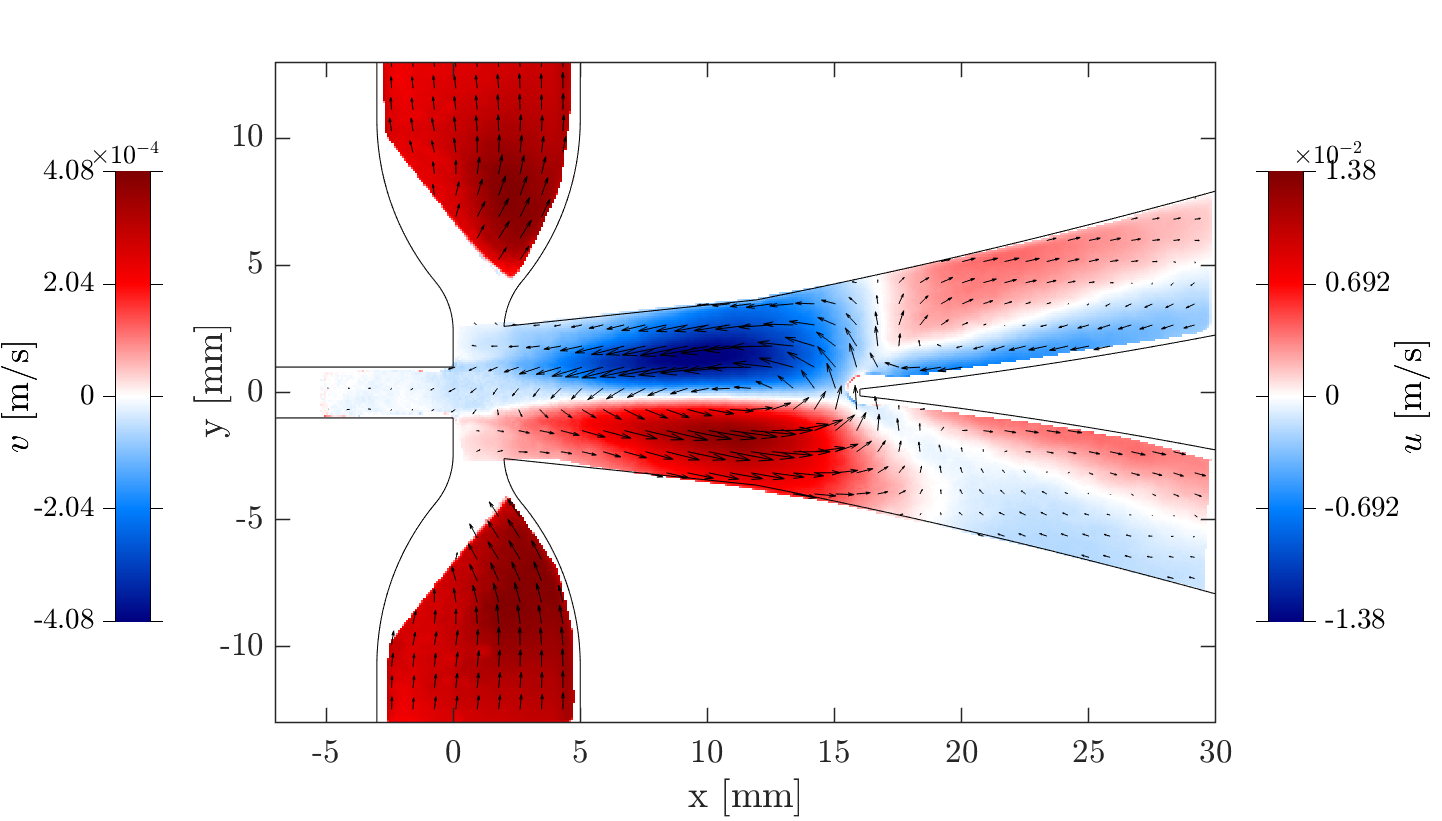}
    \caption{\label{fig:POD mode 1}Mode 1: Sweeping Mode, $\mathbf{u}_1$}
    \end{subfigure}
    \hfill
    \begin{subfigure}{0.49\textwidth}
    \includegraphics[width=\textwidth]{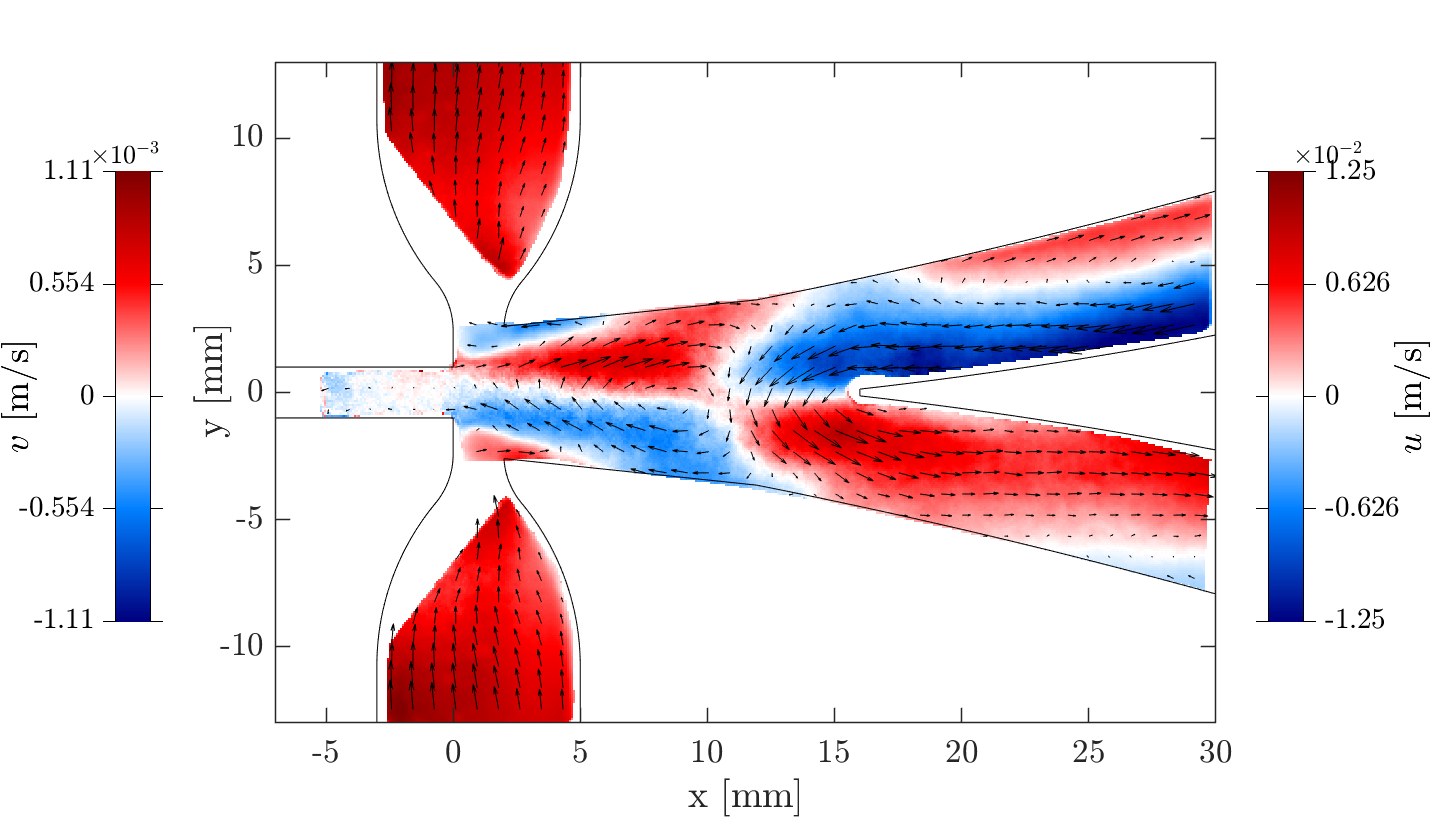}
    \caption{\label{fig:POD mode 2}Mode 2: Bending Mode, $\mathbf{u}_2$}
    \end{subfigure}
    \vspace{0.5em}
    \begin{subfigure}{0.49\textwidth}
    \includegraphics[width=\textwidth]{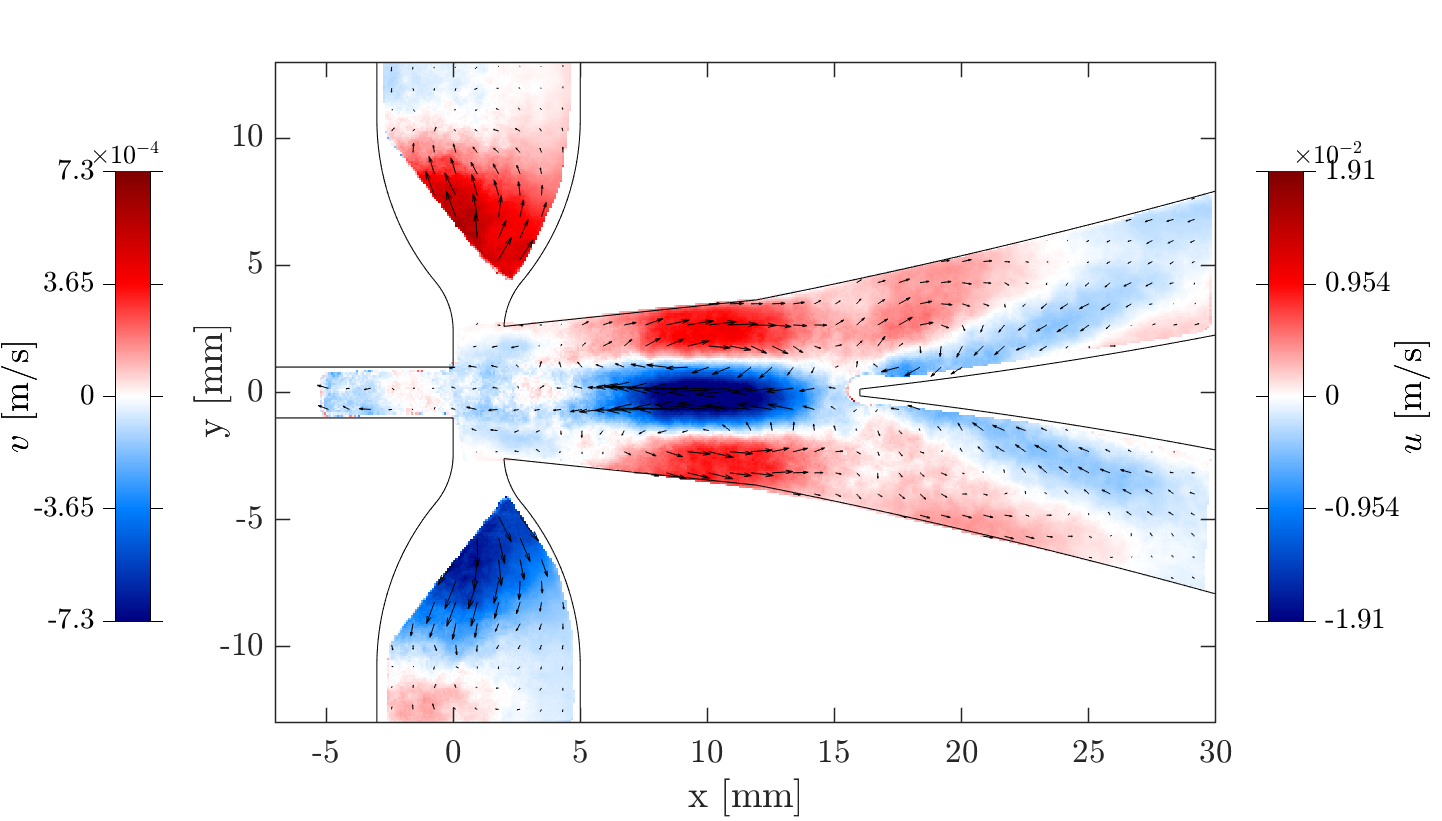}
    \caption{\label{fig:POD mode 3}Mode 3: Spreading Mode A, $\mathbf{u}_3$}
    \end{subfigure}
    \hfill
    \begin{subfigure}{0.49\textwidth}
    \includegraphics[width=\textwidth]{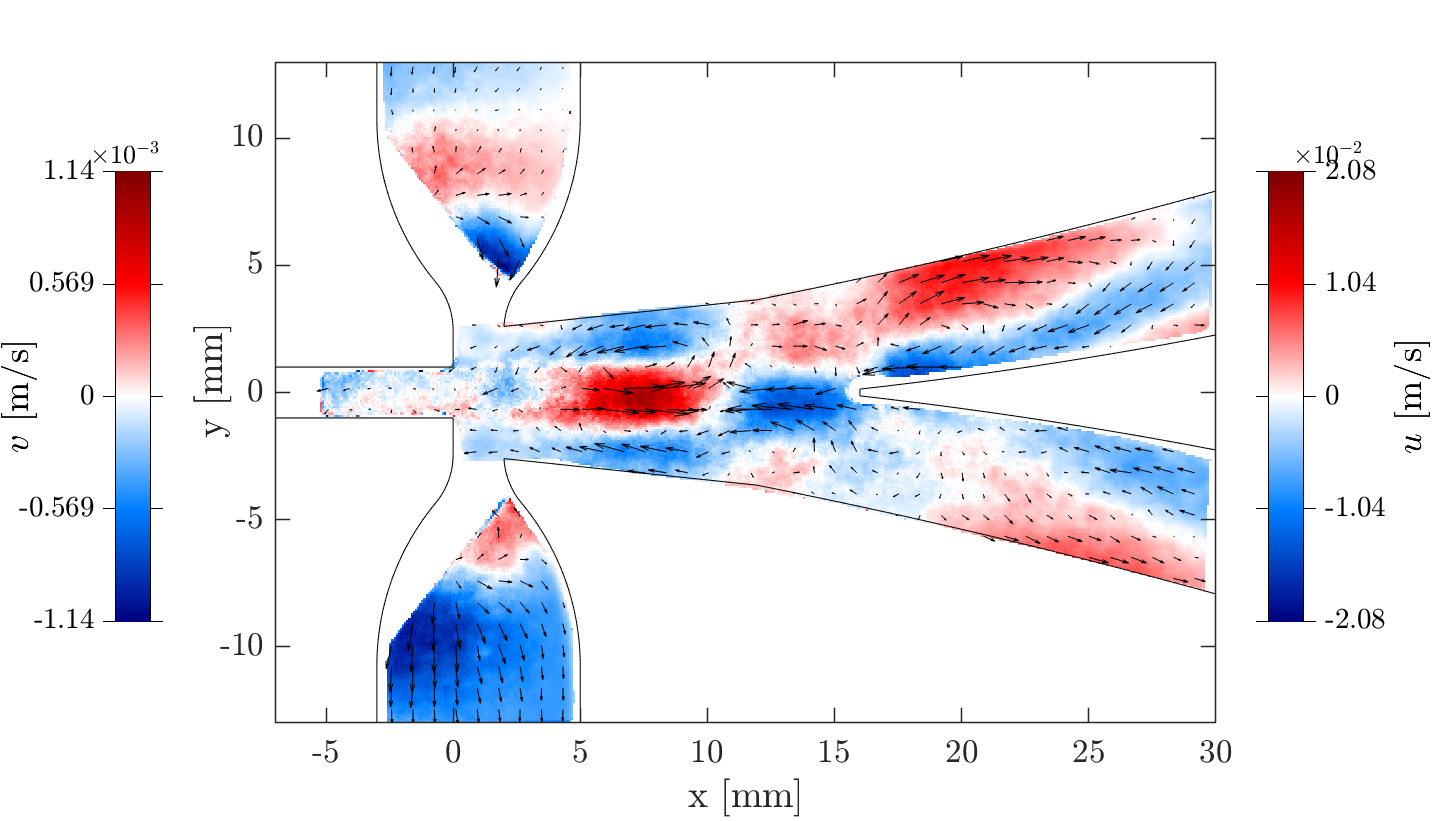}
    \caption{\label{fig:POD mode 4}Mode 4: Spreading Mode B, $\mathbf{u}_4$}
    \end{subfigure}
    \caption{Leading POD modes.}
    \label{fig:POD modes}
\end{figure}

\begin{figure}
    \centering
    \includegraphics[width=0.75\textwidth]{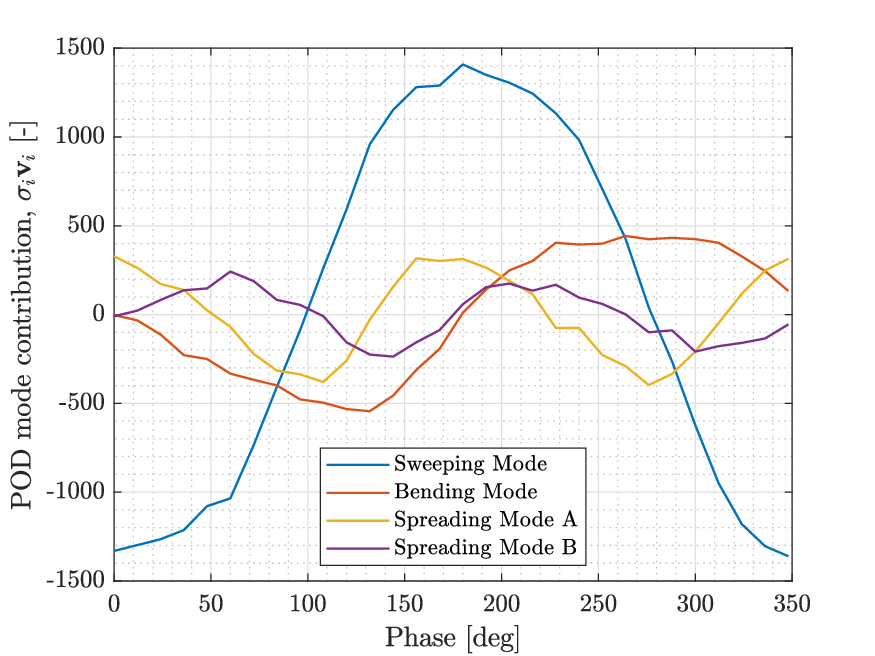}
    \caption{POD mode temporal coefficients vs phase.}
    \label{fig:POD temporal coefficients vs phase}
\end{figure}

A proper orthogonal decomposition (POD) analysis was conducted to identify the
coherent flow structures governing (i) jet motion, (ii) detachment/reattachment,
and (iii) outlet mass flux modulation. Space-only POD was applied to the
phase-averaged data via the method of snapshots using the singular value
decomposition (SVD) \citep{sirovich1987turbulence,berkooz1993proper}. The
$u$- and $v$-velocity components from both regions were stacked into a single
data matrix $\mathbf{X} \in \mathbb{R}^{2N \times M}$, whose mean was removed
prior to computing the SVD:
\begin{equation}
\mathbf{X} = \begin{bmatrix} \mathbf{u}_1 & \cdots & \mathbf{u}_M  \\
                              \mathbf{v}_1 & \cdots & \mathbf{v}_M \end{bmatrix},
\quad \widetilde{\mathbf{X}} = \mathbf{X} - \overline{\mathbf{X}},
\quad \widetilde{\mathbf{X}} = \mathbf{U}\mathbf{\Sigma}\mathbf{V}^\top
\end{equation}
where $\overline{\mathbf{X}} \in \mathbb{R}^{2N}$ is the average over the $M$ phases ($M=30$) and $N$ is the number of points where the velocity is measured. Figure \ref{fig:POD singular cumulative} shows that around
97\% of the signal variance is captured by only four POD modes, reflecting the
compact dynamics characteristic of oscillator flows \citep{noack2003hierarchy}.
The leading four modes are shown in Fig.\,\ref{fig:POD modes} and referred to
as the Sweeping Mode, Bending Mode, and Spreading Modes A \& B, corresponding
to lateral translation, curvature change, and two complementary spreading
corrections respectively. The temporal coefficients $\sigma_i\mathbf{v}_i$ are
plotted against phase in Fig.\,\ref{fig:POD temporal coefficients vs phase}.

The Sweeping Mode (Mode 1) is dominated by a large elongated vortex filling the
interaction region between the nozzle orifice and the splitter tip. Its sinusoidal
contribution at the fundamental frequency accounts for 81\% of the signal variance
and produces a sweeping jet motion when combined with the mean flow. Its
contribution to outlet channel mass flux modulation is small; it primarily
describes lateral jet position rather than outlet flow split. The feedback channel region
of the Sweeping Mode supplies the entrainment flow into the recirculation bubble
in phase with the main jet position.

The Bending Mode (Mode 2) describes jet curvature around a pivot at
$x \approx 11$\,mm, $y = 0$\,mm, accounting for around 10\% of the signal
variance. Its contribution $\sigma_2\mathbf{v}_2$ is principally at the fundamental frequency, lagging the Sweeping Mode by $90^\text{o}$, and it is associated with detachment and reattachment. At $\phi = 0^\text{o}$ the rate of change of $\sigma_2\mathbf{v}_2$ is large relative to $\sigma_1\mathbf{v}_1$, so detachment begins with a curvature change rather than a lateral translation. Throughout the detachment process, both modes contribute a down-wash through the feedback channel that expands
the recirculation bubble; the Bending Mode describes how the jet bends around the expanding recirculation bubble.
At $\phi = 100^\text{o}$ when the Sweeping
Mode contribution reaches zero, it is the Bending Mode --- sustained by its
$90^\text{o}$ phase lag --- that continues to drive bubble expansion and push
the jet towards the lower wall. The process of attachment to the lower wall
occurs between $\phi = 100^\text{o}$ and $180^\text{o}$, with the Bending Mode
reaching maximum curvature towards the lower wall at $\phi = 132^\text{o}$.
Together, the Sweeping and Bending modes form a minimal oscillator: one
describing jet position, the other its phase-shifted curvature response.

The Spreading Modes A and B (Modes 3 and 4) account for approximately $4.5\%$
and $1.5\%$ of the signal variance respectively and contribute at the first
harmonic of the oscillation frequency in quadrature
(Fig.\,\ref{fig:POD temporal coefficients vs phase}). Together they produce a
downstream-travelling pulsing correction that modulates jet spreading during
the traverse (Spreading Mode A) and as a function of streamwise position during
wall attachment (Spreading Mode B), interpreted as corrections to the spreading
and entrainment flows produced by the Sweeping and Bending Modes.

\begin{figure}
    \centering
    \includegraphics[width=0.75\textwidth]{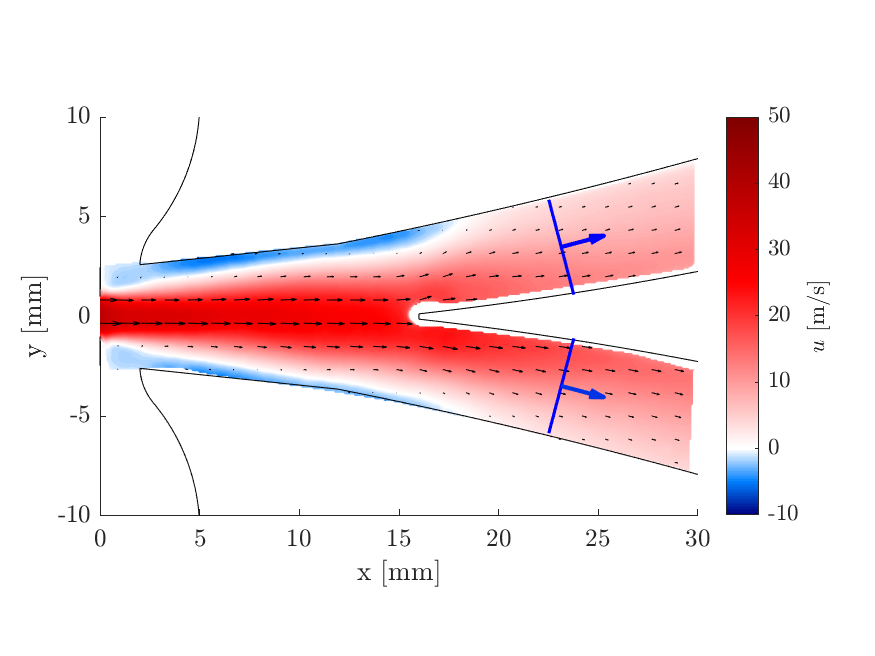}
    \caption{Mean flow with lines showing location for estimating outlet channel
    volume fluxes. The outlet restrictions are further downstream than the visible region (see Fig.\,\ref{fig:device geometries})}
    \label{fig:mean flow with outlet mass flux slices}
\end{figure}

\begin{figure}
    \centering
    \includegraphics[width=0.75\textwidth]{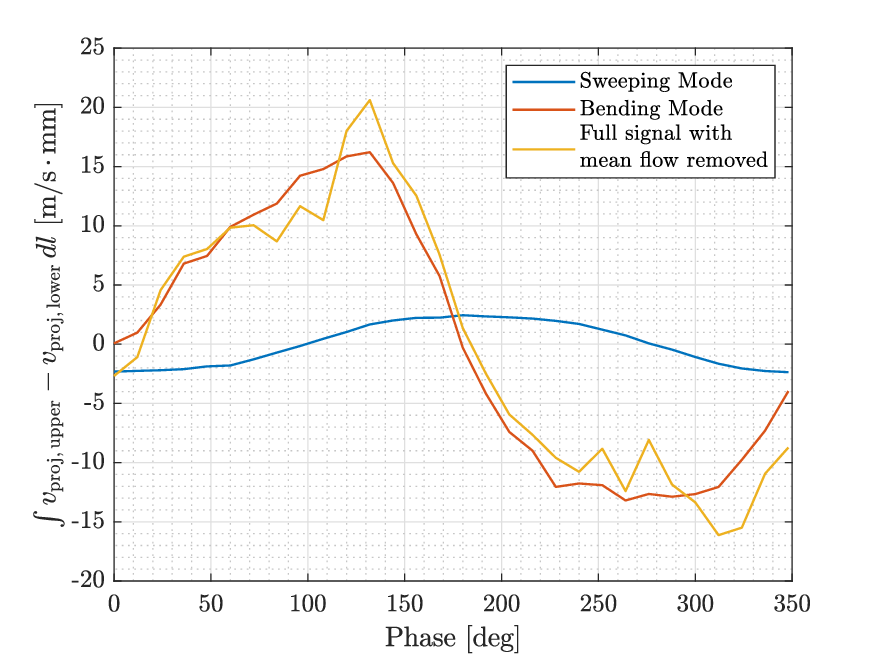}
    \caption{Differential outlet channel volume flux per unit depth (upper $-$
    lower channel) vs phase: Sweeping Mode (blue), Bending Mode (red), and full
    flow without mean (yellow).}
    \label{fig:Differential outlet channel 2D efflux vs phase}
\end{figure}

Fig.\,\ref{fig:mean flow with outlet mass flux slices} shows slices across
the mid-point of the visible outlet channels superimposed on the mean flow.
The flow field was reconstructed from the Sweeping and Bending Modes separately
($\mathbf{u}_1\sigma_1\mathbf{v}_1$ and $\mathbf{u}_2\sigma_2\mathbf{v}_2$),
projected onto the local channel directions, and averaged along each slice to
give the outlet channel volume flux per unit depth. The differential flux
(upper minus lower channel) is shown in
Fig.\,\ref{fig:Differential outlet channel 2D efflux vs phase}.
The Bending Mode makes the dominant contribution\footnote{It is shown in
Section\,\ref{sec:Effect of varying outlet restriction} that for cases where the
differential outlet volume flux amplitude is small, the Bending Mode spatial
shape changes accordingly in spite of its contribution $\sigma_2\mathbf{v}_2$
remaining significant. The Bending Mode therefore maintains an important role
in detachment and reattachment even when it does not modulate the outlet volume
flux.}. Its mode shape $\mathbf{u}_2$ introduces curvature upstream of the
splitter to bias the jet towards one outlet by adding forward flow and reverse
flow contributions on opposite sides. The outlet flow split therefore lags the
jet position by approximately $90^\text{o}$, in contrast to a conventional
fluidic diverter where the two are expected to be in phase
\citep{joyce1983fluidics}, and to a sweeping jet actuator where they are in
anti-phase \citep{gokoglu2009numerical}. This $90^\text{o}$ lag arises from the
tendency of the jet to attach to the splitter, which weakens the link between
upstream jet position and outlet flow split --- a feature explored further in
Section\,\ref{sec:Effect of varying outlet restriction}.

The POD analysis reveals that the oscillator dynamics are governed by two coupled
degrees of freedom: lateral jet displacement (Sweeping Mode) and curvature-induced
detachment (Bending Mode). Crucially, outlet mass flux modulation is determined
by jet bending rather than lateral displacement, weakening the direct link between
upstream jet attachment and outlet flow split.
\subsection{Circulation in the feedback channel entrances}
\label{sec:Circulation in control ports}
The POD analysis in Section\,\ref{sec:POD analysis} revealed that the feedback channel flow plays a central role in the oscillation, supplying the mass flux that aspirates the separation bubbles and drives jet switching. The mean flow in Fig.\,\ref{fig:mean flow} shows a time-mean recirculating vortex at each feedback channel entrance, confirmed to be driven by the time-mean jet rather than the oscillatory component by the negligible circulation in the POD modes (Fig.\,\ref{fig:POD modes}). This recirculation reduces the effective flow area at the channel entrances, thereby limiting the mass flux available to aspirate the separation bubbles and drive switching.

\begin{figure}[h]
\centering
\begin{subfigure}{0.40\textwidth}
\includegraphics[width=\textwidth]{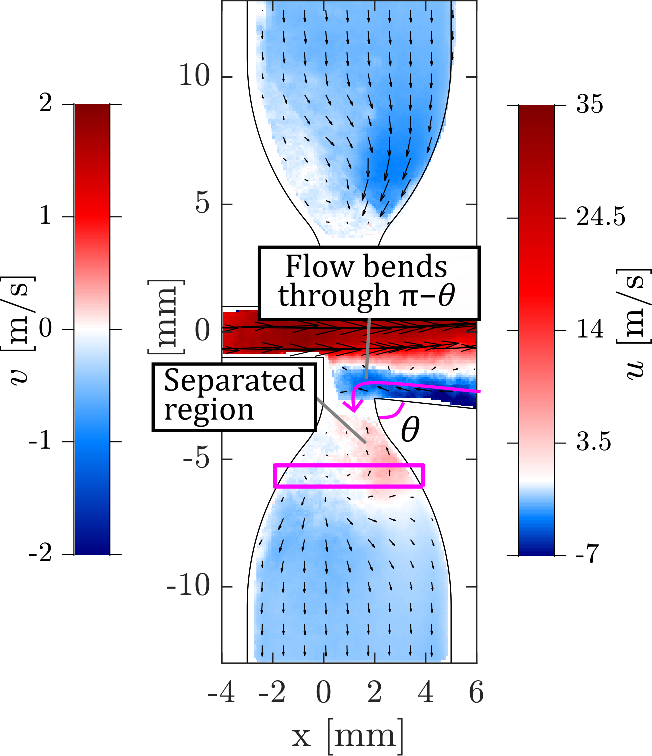}
\caption{Extraction region (pink box)\label{fig:v-vel profile flow field}}
\end{subfigure}
\hfill
\begin{subfigure}{0.57\textwidth}
\includegraphics[width=\textwidth]{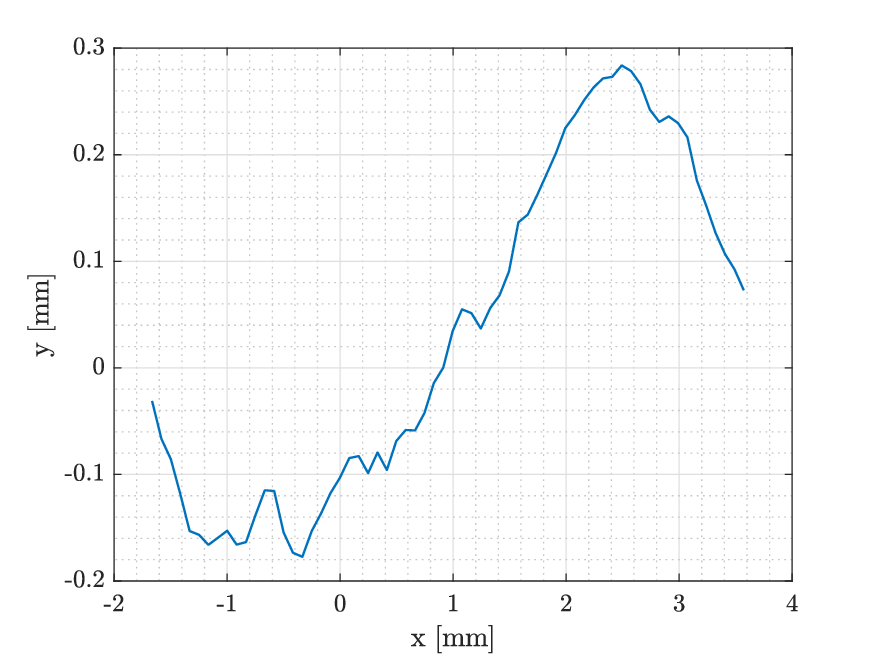}
\caption{$v$-velocity profile \label{fig:v-vel profile curve}}
\end{subfigure}
\caption{$v$-velocity extraction at $\phi=36^\text{o}$, $\dot{m}=0.49$\,g/s:
(a) extraction region; (b) resulting profile.}
\label{fig:v-velocity profile across contracted region}
\end{figure}

Fig.\,\ref{fig:v-velocity profile across contracted region} shows data from $\dot{m} = 0.49$\,g/s, where the PIV domain at the control port is more complete. The separated region on the right-hand side of the lower feedback channel is highlighted in Fig.\,\ref{fig:v-velocity profile across contracted region}(a), with the corresponding $v$-velocity profile --- averaged across $y = -6.1$ to $-5.2$\,mm at $\phi = 36^\text{o}$ --- shown in Fig.\,\ref{fig:v-velocity profile across contracted region}(b). The profile confirms that the down-flow is restricted to the left-hand side of the channel, reducing its effective width. The separation arises because the recirculating flow must turn through a large angle ($\pi - \theta$ in Fig.\,\ref{fig:v-velocity profile across contracted region}(a)) to enter the channel, a consequence of the channel expanding from the connection point to its full width. While the extent of this separation is most probably exaggerated in the present geometry, an analogous effect is expected in any perpendicular feedback channel connection where the flow must undergo a sharp turn at entry.

\begin{figure}[h]
\centering
	\begin{subfigure}{0.38\textwidth}
	\includegraphics[width=\textwidth]{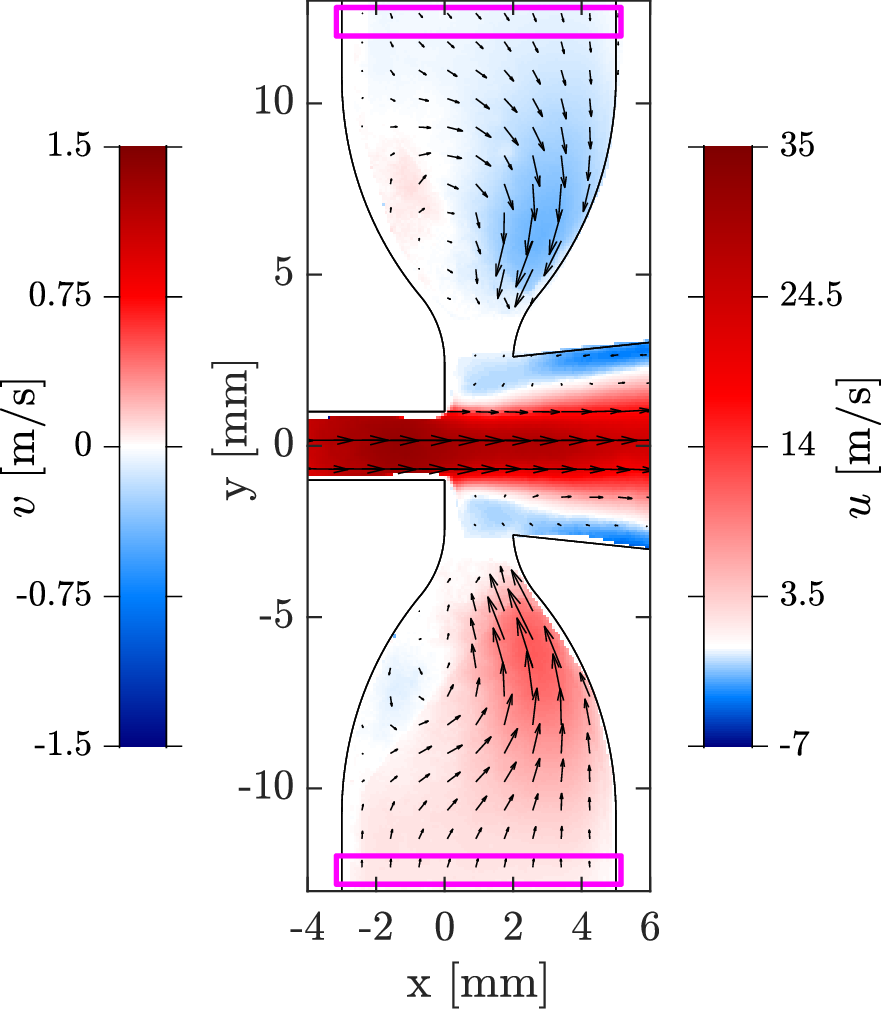}
	\caption{\label{fig:23slpm mean flow zoom on FC with slices}Mean flow at $0.49$\,g/s}
	\end{subfigure}
	\begin{subfigure}{0.6\textwidth}
	\includegraphics[width=\textwidth]{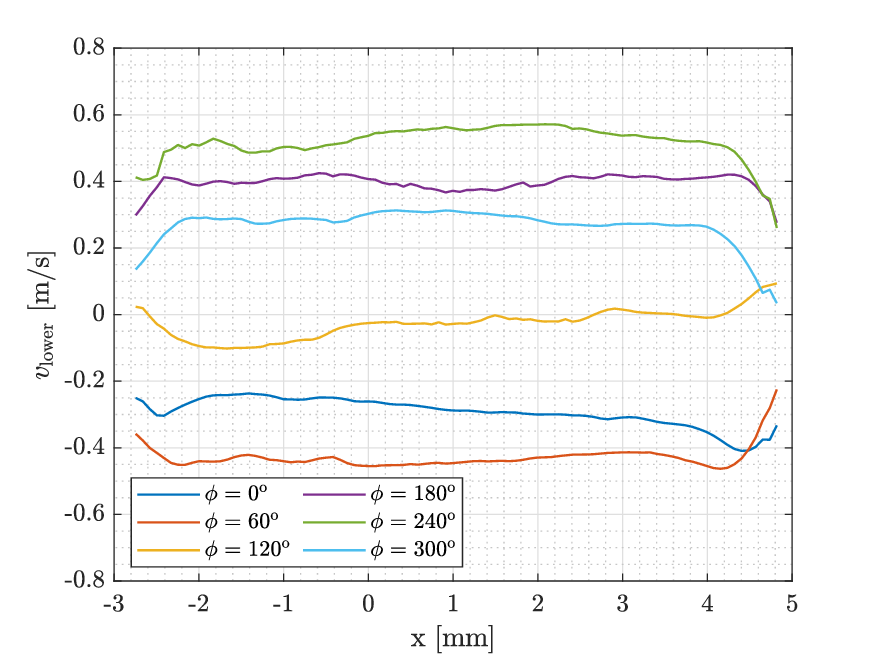}
	\caption{\label{fig:23slpm mean flow FC velocity profile}$v$-velocity profiles at $0.49$\,g/s at several phases}
	\end{subfigure}
	\begin{subfigure}{0.38\textwidth}
	\includegraphics[width=\textwidth]{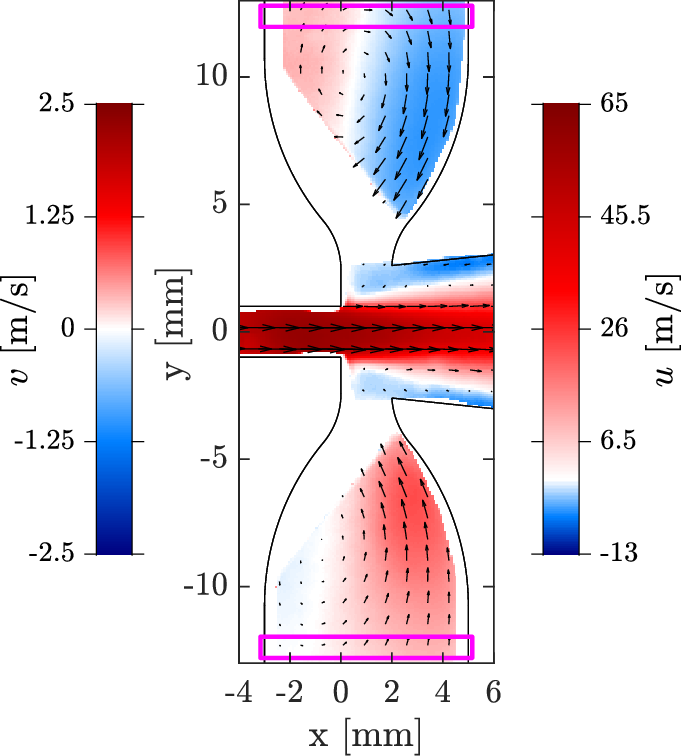}
	\caption{\label{fig:43slpm mean flow zoom on FC with slices}Mean flow at $0.91$\,g/s}
	\end{subfigure}
	\begin{subfigure}{0.6\textwidth}
	\includegraphics[width=\textwidth]{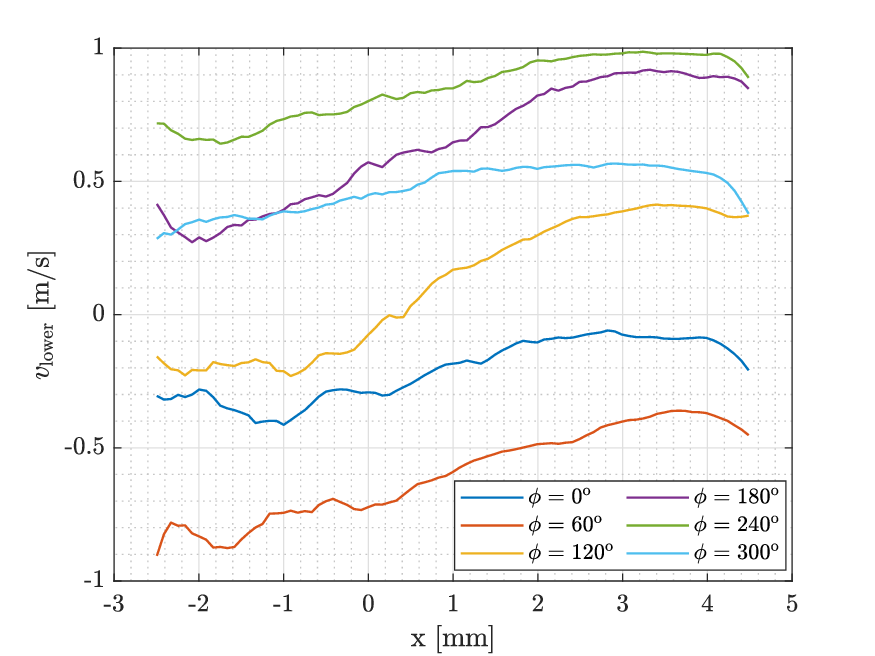}
	\caption{\label{fig:43slpm mean flow FC velocity profile}$v$-velocity profiles at $0.91$\,g/s at several phases}
	\end{subfigure}
	\caption{Mean flow fields at $0.49$\,g/s (a) and $0.91$\,g/s (c) showing slices for extracting velocity profiles. Corresponding velocity profiles shown for $0.49$\,g/s (b) and $0.91$\,g/s (d).}
	\label{fig:mean flows and velocity profiles vs phase}
\end{figure}

Fig.\,\ref{fig:mean flows and velocity profiles vs phase} shows the mean flow zoomed on the feedback channels for $\dot{m} = 0.49$\,g/s and $0.91$\,g/s, with $v$-velocity profiles extracted across the channel at the furthest visible point from the interaction region shown alongside each mean flow in Fig.\,\ref{fig:mean flows and velocity profiles vs phase}(b,d). At $0.49$\,g/s the profiles are approximately uniform, while at $0.91$\,g/s a significant recirculation is evident from the diagonal shape of the profiles in Fig.\,\ref{fig:43slpm mean flow FC velocity profile}. At the higher flow rate, the separation bubbles also extend further into the channel. Together, these observations indicate a stronger recirculation at higher flow rates and hence a more pronounced throttling of the feedback flow.

\begin{figure}
\centering
\includegraphics[width=0.6\textwidth]{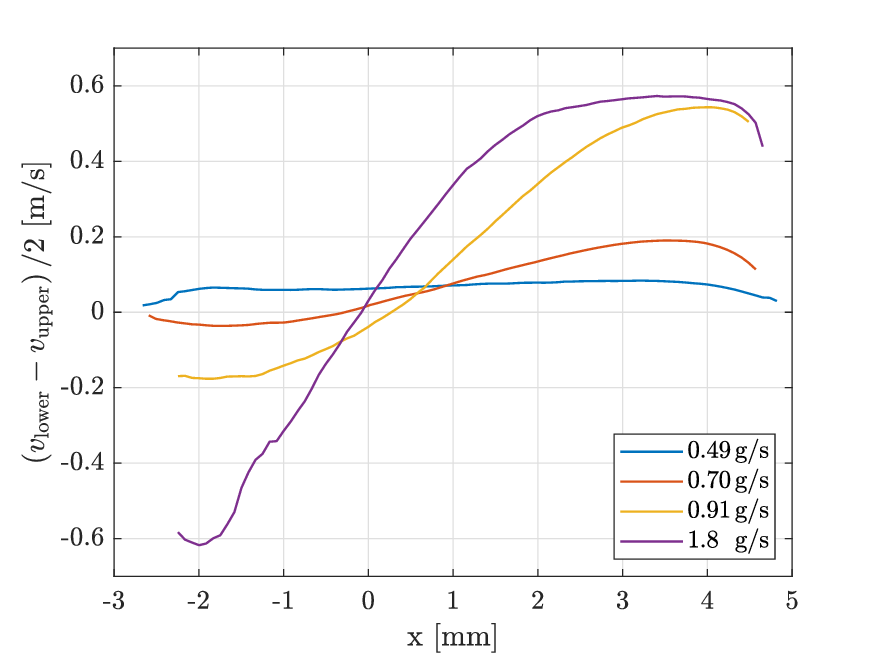}
\caption{\label{fig:time mean v-velocity profiles averaged across upper and lower FCs}Time-mean $v$-velocity profiles combined between upper and lower feedback channels ($\left(v_\text{lower}-v_\text{upper}\right)/2$) at several inlet mass flow rates.}
\end{figure}

Fig.\,\ref{fig:time mean v-velocity profiles averaged across upper and lower FCs} shows time-mean $v$-velocity profiles across the feedback channel entrances for several inlet flow rates, combined as $(\bar{v}_\text{lower} - \bar{v}_\text{upper})/2$ to account for the opposing circulation directions in the upper and lower channels. With increasing flow rate, the recirculation extends further from the interaction region. The local Reynolds number at the entrances is $Re \sim O(100)$, a regime sensitive to flow rate. This is consistent with a progressively stronger throttling of the feedback flow as the inlet flow rate increases and scales the local Reynolds number.

\begin{figure}
    \centering    
    \includegraphics[width=0.65\textwidth]{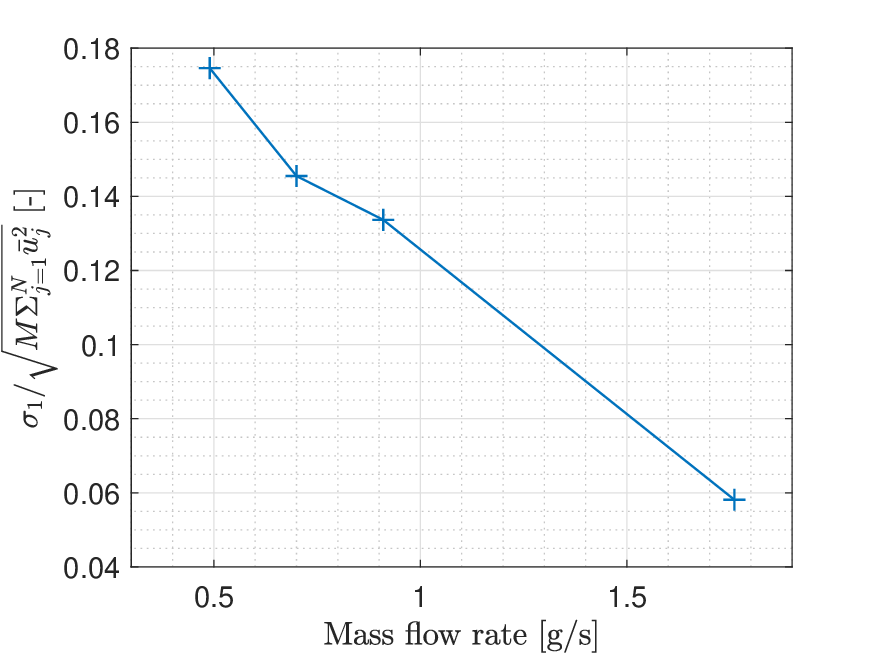}
    \caption{Singular value of Sweeping Mode normalised by RMS of mean flow field vs inlet mass flow rate}
    \label{fig:mode 1 SV vs mass flow rate.}
\end{figure}

Fig.\,\ref{fig:mode 1 SV vs mass flow rate.} shows the Sweeping Mode singular value normalised by the mean flow RMS, $\sigma_1/\sqrt{M\sum_{j=1}^N\left[\bar{\mathbf{u}} \;\bar{\mathbf{v}}\right]^\top}$, plotted against inlet mass flow rate. The decreasing trend confirms that oscillation strength weakens with increasing flow rate, consistent with the progressive throttling of the feedback flow by the growing recirculation at the channel entrances. Although exaggerated in the present device by the particular geometry of the feedback channel connection, this effect is expected to persist in any perpendicular configuration since the flow must still undergo a sharp turn at entry. The aerodynamic design of the feedback channel connection is therefore an important factor in oscillator performance, despite the low local flow speeds.

In the following section the effect of outlet restriction is examined, which reveals a contrasting behaviour: whereas increasing feedback channel impedance weakens the oscillation, increasing outlet restriction does not suppress it --- a result that challenges the conventional understanding of how back pressure influences jet attachment.
\subsection{Effect of varying outlet restriction}
\label{sec:Effect of varying outlet restriction}
One point this paper aims to address is how the differential back pressure induced by restrictive outlet flow paths, such as those studied by \citet{guyot2008active} and \citet{nicholls2022novel}, influences the operation of a sonic oscillator. An uneven outlet mass flux split induces a different recovery of dynamic pressure in each outlet channel, producing a differential back pressure that is transmitted upstream through the outlet channels and acts across the jet at the splitter. This pressure difference tends to peel the jet off the wall from downstream of the reattachment point and counteracts the Coand\u{a} effect. The feedback channel flow also acts to destabilise the Coand\u{a} effect through the aspiration of the recirculation bubble. However, the steady state flow through the feedback channel may not exceed the switching threshold. In this case, the jet remains stably attached to one of the walls. The introduction of outlet restriction weakens the jet attachment from downstream and can instigate the oscillation.

If the outlet aperture width, $w_\text{o}$, is reduced beyond the onset of the oscillation, the increasing differential back pressure is expected to limit the extremes of the jet position by reducing the strength of the Coand\u{a} effect. This in turn would limit the modulation of the outlet mass flux split, thereby moderating the differential back pressure. As $w_\text{o}$ is reduced further, the increasing back pressure should suppress jet attachment entirely, with the jet flowing symmetrically down the centreline in the limit $w_\text{o} \to 0$. Under this hypothesised behaviour, the upstream jet attachment strength and outlet mass flux split remain directly linked across all values of $w_\text{o}$. As demonstrated below, the observed behaviour departs significantly from this expectation.

\begin{figure}[h]
    \centering    
    \includegraphics[width=0.65\textwidth]{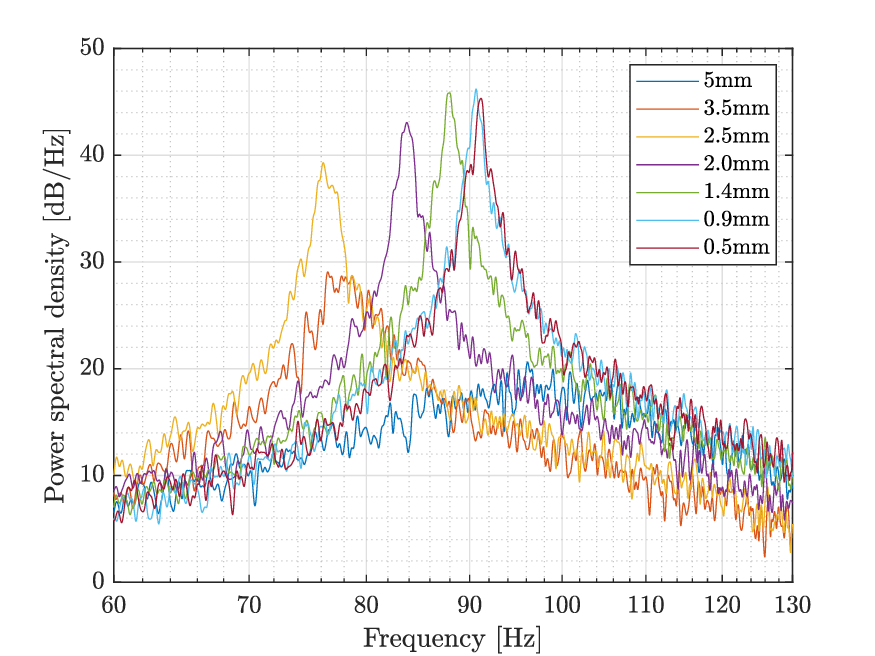}
    \caption{Power spectral density of differential outlet channel pressure signals at several outlet apertures.}
    \label{fig:pressure spectra at different outlet restrictions}
\end{figure}

Experiments were conducted at outlet apertures in the range $0.5 \leq w_\text{o} \leq 5$\,mm (contraction ratios from 14.4 to 1.4). Figure \ref{fig:pressure spectra at different outlet restrictions} shows the power spectral density of the differential outlet channel pressure signal for each case. The spectra were computed from $2.15\times10^6$ points sampled at $50$\,kHz using Welch's method with 10 segments, 50\% overlap, and a Hamming window, and corrected for the frequency response of the transducers and pneumatic connections using a planar wave tube and a flat-response reference microphone. The reference pressure is $1$\,Pa.

At $w_\text{o} = 5$\,mm the oscillation peak is broad and at a relatively high frequency, reflecting the large phase noise when the jet remains attached to one wall throughout and natural fluctuations dominate the weak limit cycle. As $w_\text{o}$ reduces from $3.5$\,mm to $0.5$\,mm the spectral peak sharpens and shifts to higher frequencies, approaching an asymptotic value at the smallest apertures rather than collapsing as hypothesised.

\begin{figure}[h]
    \centering    
    \includegraphics[width=0.85\textwidth]{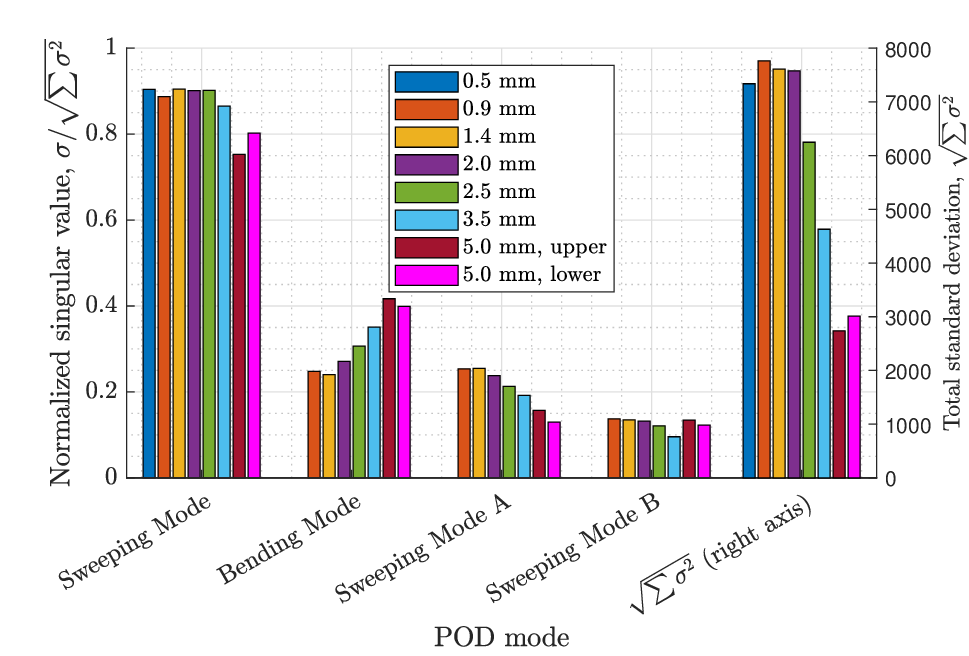}
    \caption{POD mode singular values normalised by signal standard deviation (left-hand y-axis) and total signal standard deviation (no mean, right-hand y-axis) for several outlet apertures.}
    \label{fig:POD mode singular values for several outlet restrictions}
\end{figure}

Fig.\,\ref{fig:POD mode singular values for several outlet restrictions} shows the POD mode singular values of the phase-averaged PIV data at each outlet aperture, normalised by the total signal standard deviation (left axis), with the total standard deviation also shown (right axis)\footnote{At $w_\text{o} = 0.5$\,mm the modal structure changes: the second and third modes merge into a pair incorporating aspects of both the Bending and Spreading Modes at both the fundamental frequency and its first harmonic, so no separate Bending or Spreading Mode data are shown for this case.}. With decreasing $w_\text{o}$, signal energy shifts from the Bending Mode to the Spreading Modes. Most significantly, the total signal variance \textit{increases} as $w_\text{o}$ is reduced while the Sweeping Mode importance remains invariant, confirming that oscillation strength grows rather than collapses as $w_\text{o} \to 0$.%

The key insight is that under these conditions, the differential outlet mass flux is not governed substantially by the oscillation strength, and the oscillation strength is only partially governed by the differential outlet mass flux --- contrary to the assumption that upstream jet attachment and outlet flow split are always coupled. The following analysis demonstrates how this coupling breaks down and explains the underlying physical mechanism.
%
\subsubsection{Phasor analysis}
\begin{figure}[h]
\centering
	\begin{subfigure}{0.49\textwidth}
	\includegraphics[width=\textwidth]{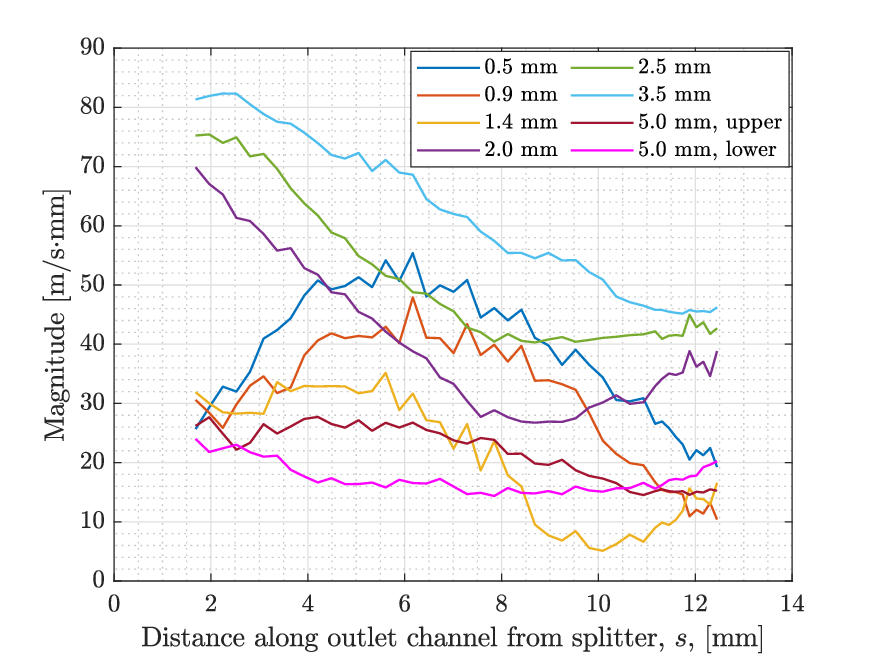}
	\caption{\label{fig:q_RMS vs several wo values}Magnitude}
	\end{subfigure}
	\begin{subfigure}{0.49\textwidth}
	\includegraphics[width=\textwidth]{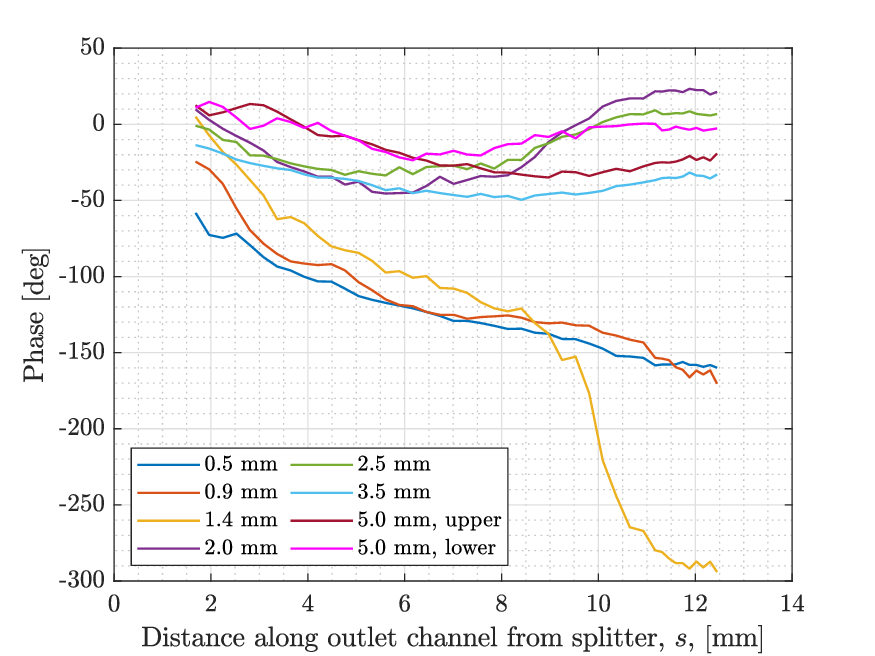}
	\caption{\label{fig:arg(q) vs several wo values}Phase}
	\end{subfigure}
	\caption{Magnitude and phase of differential outlet channel volume flux per unit depth at fundamental oscillation frequency, $q(s)$, vs distance along outlet channel, $s$.}
	\label{fig:magnitude and phase of differential volume flux}
\end{figure}

Although the spatial POD modes $\mathbf{u}_i$ change somewhat with $w_\text{o}$, their qualitative interpretation remains consistent. Of particular interest is the contribution of each mode to the differential outlet volume flux, which determines the differential back pressure via dynamic pressure recovery at the restrictions. The flow fields were reconstructed from the Sweeping and Bending Modes\footnote{The first three modes were used for $w_\text{o}=0.5$\,mm owing to its different modal structure.} and the volume flux per unit depth computed across several slices along the outlet channels, following the method of Section\,\ref{sec:POD analysis}. A phasor representation describes the magnitude and phase of the volume flux at the fundamental frequency at each streamwise location, $q_\text{U}(s), q_\text{L}(s) \in \mathbb{C}$, and the differential phasor $q_{\Delta}(s) = q_\text{U}(s) - q_\text{L}(s)$ is plotted against distance along the channel centreline from the splitter tip, $s$, in Fig.\,\ref{fig:magnitude and phase of differential volume flux}.

For a purely two-dimensional incompressible flow the volume flux per unit depth would be constant along the channel; the variation of $|q_\Delta(s)|$ with distance therefore indicates either a spanwise flow component or compressibility such as acoustic wave propagation. At $w_\text{o} = 5$\,mm the magnitude is small throughout because the jet remains attached to one wall, heavily biasing the time-averaged outlet flow and reducing the variance (Fig.\,\ref{fig:POD mode singular values for several outlet restrictions}). For $w_\text{o} \geq 2.0$\,mm the phase of $q_\Delta(s)$ is approximately constant along the channel, while for $w_\text{o} \leq 1.4$\,mm it decreases monotonically, indicating the presence of a convecting structure superposed on the spatially-averaged component.
The wavelength of an acoustic wave at the oscillation frequencies in the present paper are in excess of 3\,m, whereas the length of the outlet channel from the splitter to the outlet restriction is 50\,mm, which suggests that the outlet channels are acoustically compact. Additionally, the gradients of the phase-distance curves for $w_\text{o} \leq 1.4$\,mm in Fig.\,\ref{fig:arg(q) vs several wo values} correspond to a convection speed of $O(1)$\,m/s. These factors suggest that the convecting structure is hydrodynamic rather than a standing acoustic wave. However, single reflections from the restriction --- which acts as a near-closed acoustic boundary --- cannot be excluded on this basis alone, as such reflections would return to the splitter on a timescale short compared with the oscillation period and would be indistinguishable from the hydrodynamic back pressure in the present measurements.
Since it is the spatially-averaged differential flux that determines the net hydrodynamic back pressure, the convective component is removed before comparing cases. First, the origin of the convecting structure is identified. 
\begin{figure}[h]
\centering
	\begin{subfigure}{0.6\textwidth}
	\includegraphics[width=\textwidth]{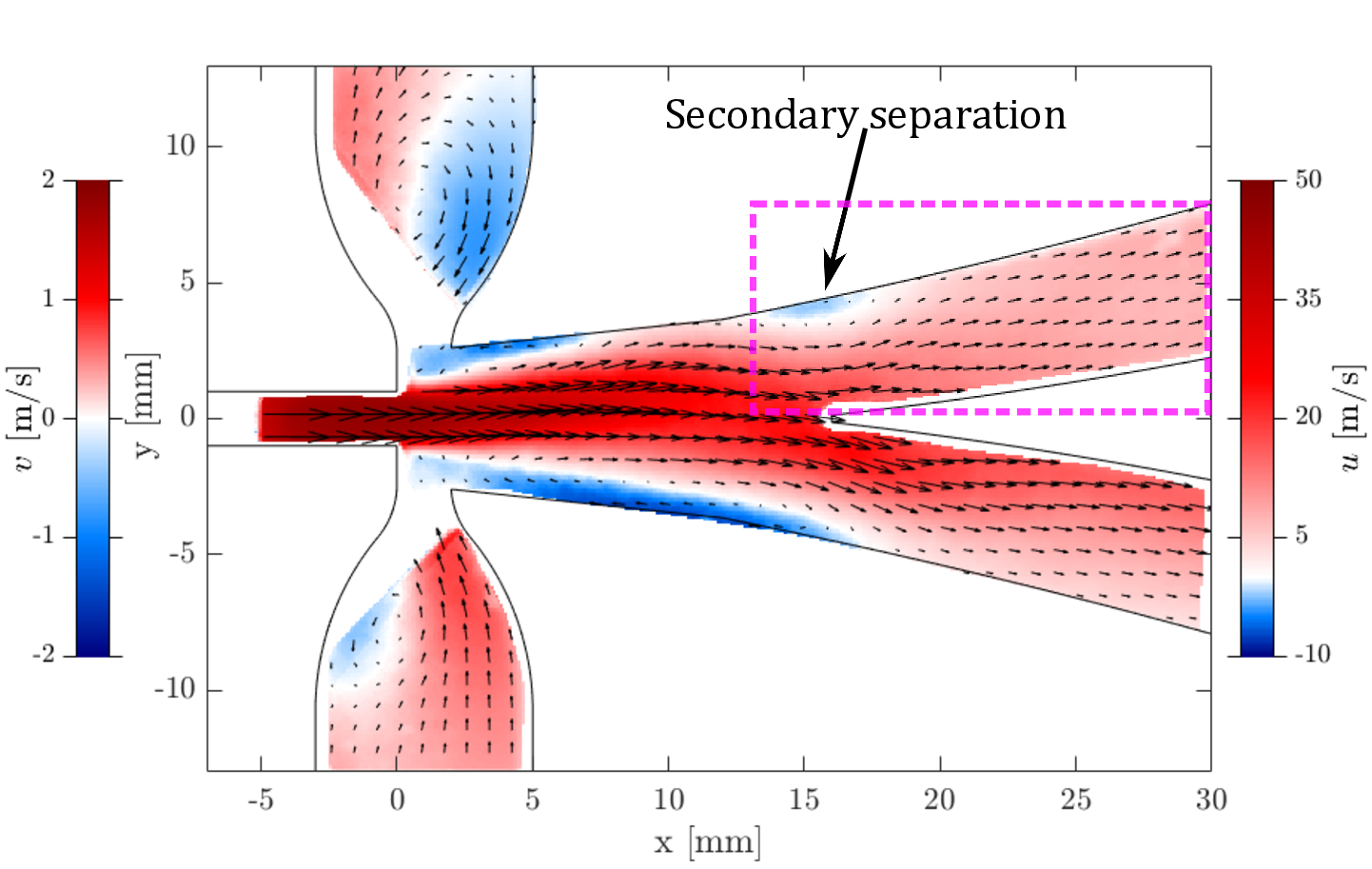}
	\caption{\label{fig:flow snapshot 1.4mm phase 25 annotated at secondary separation}Flow snapshot at $\phi=288^\text{o}$, showing location of secondary separation in upper outlet channel}
	\end{subfigure}
	%
	%
	\begin{subfigure}{0.5\textwidth}
	\includegraphics[width=\textwidth]{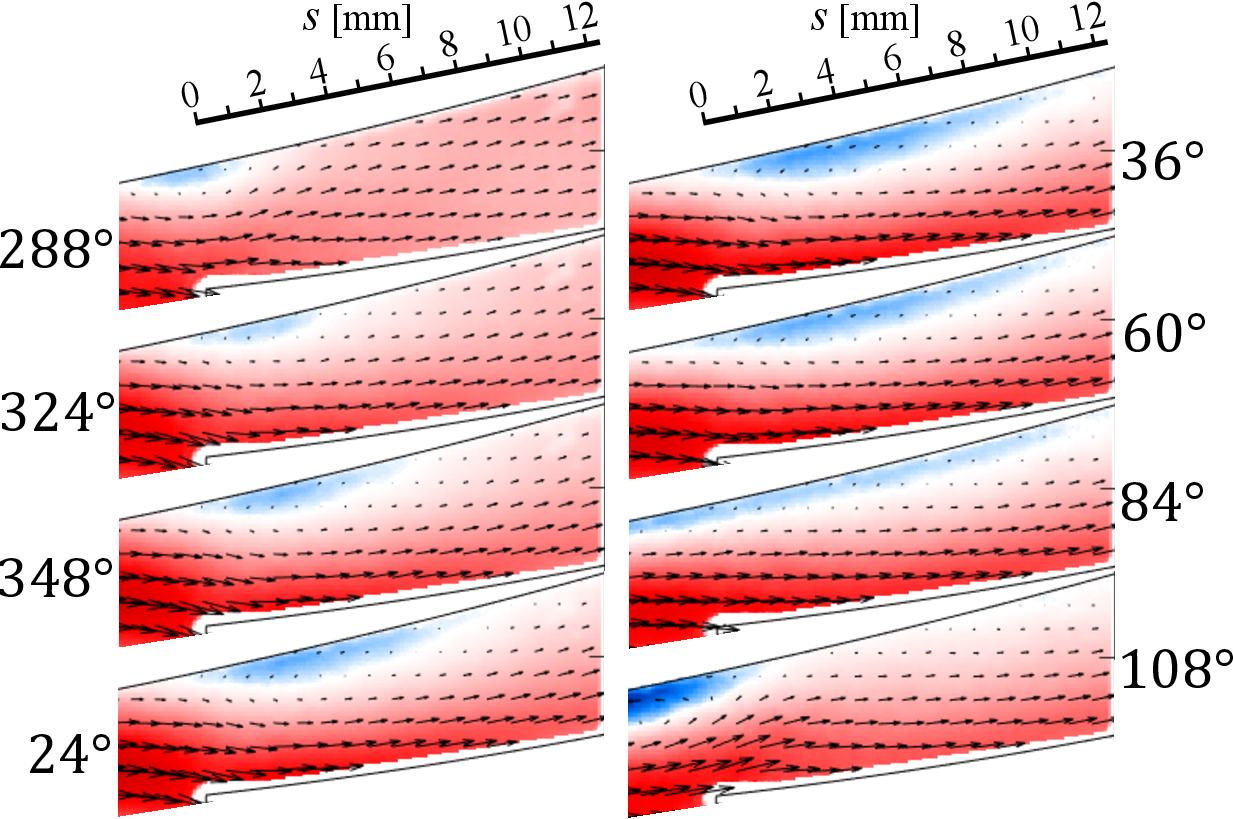}
	\caption{\label{fig:outlet channel zoom over several phases}Downstream growth and subsequent collapse of secondary separation bubble over several phases. Scale, $s$, indicates distance along channel centreline with the splitter tip as the datum.}
	\end{subfigure}
	\caption{Outlet channel secondary separation bubble at $w_\text{o} =1.4$\,mm}
	\label{fig:Outlet channel secondary separation bubble at 1.4mm}
\end{figure}

Fig.\,\ref{fig:flow snapshot 1.4mm phase 25 annotated at secondary separation} shows a phase snapshot of the flow at $w_\text{o}=1.4$\,mm and $\phi=288^\text{o}$. The secondary separation in the upper outlet channel is highlighted, and Fig.\,\ref{fig:outlet channel zoom over several phases} shows the region of the flow in the pink box over several phases that follow. The secondary separation can be seen to move downstream and grow in size before eventually collapsing. This structure is believed to be responsible for the streamwise variation in volume flux observed in Fig.\,\ref{fig:arg(q) vs several wo values}.
\begin{figure}[h]
\centering
    \begin{subfigure}{0.6\textwidth}
    \includegraphics[width=\textwidth]{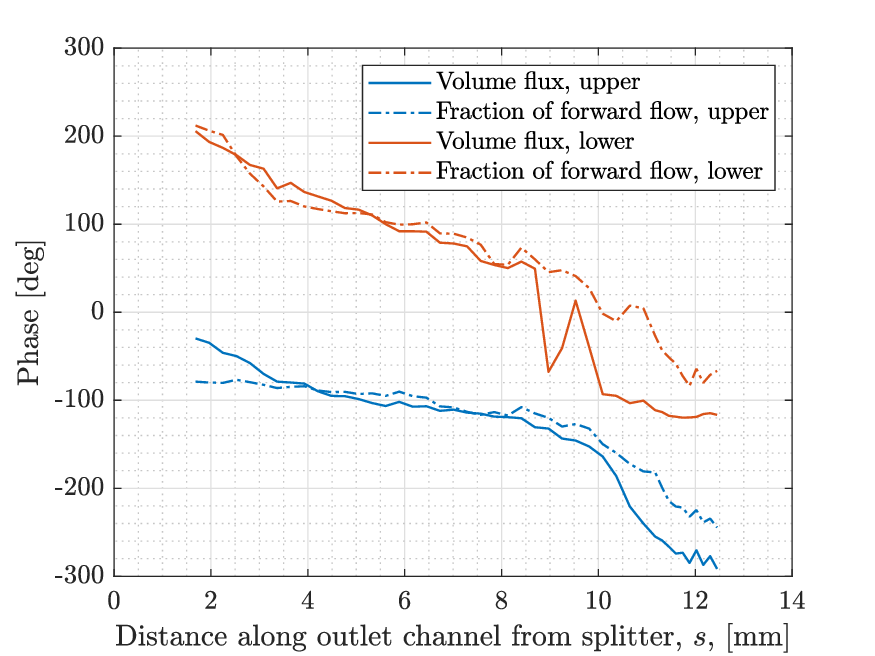}
    \caption{\label{fig:phase of phasors of qFlux and fraction of forward flow}
    Phase vs distance along outlet channel.}
    \end{subfigure}
    \vspace{0.5em}
    \begin{subfigure}{0.49\textwidth}
    \includegraphics[width=\textwidth]{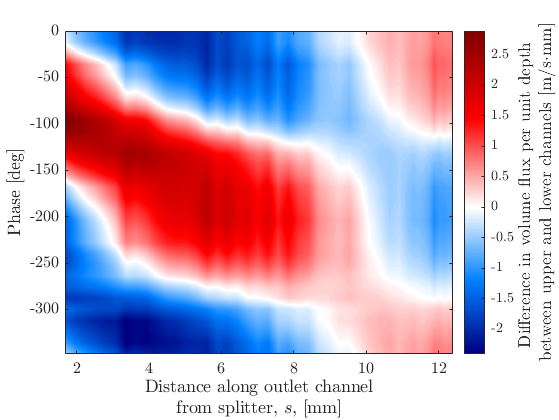}
    \caption{\label{fig:Phase-space plot of difference in volume flux}
    Volume flux per unit depth.}
    \end{subfigure}
    \hfill
    \begin{subfigure}{0.49\textwidth}
    \includegraphics[width=\textwidth]{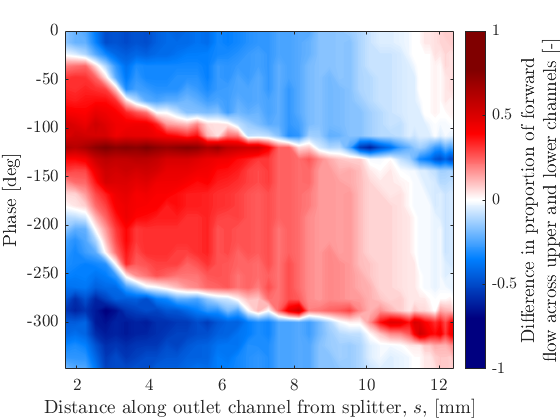}
    \caption{\label{fig:Phase-space plot of difference in fractional width of forward flow}
    Fractional width of forward flow.}
    \end{subfigure}
    \caption{Comparison between volume flux per unit depth and fractional width 
    of forward flow at $w_\text{o}=1.4$\,mm. (a) Phase of phasor at the 
    fundamental oscillation frequency vs distance along the outlet channel, $s$ for 
    volume flux per unit depth (solid lines) and fractional width of forward flow 
    (dash-dot lines) in the upper (blue) and lower (red) outlet channels. 
    (b,c) Phase-space plots of the difference in (b) volume flux 
    per unit depth and (c) fractional width of forward flow.}
    \label{fig:comparison between volume flux and fractional forward flow phasors and phase-space plots}
\end{figure}

To link the growth and decay of the secondary separation bubble to the volume flux, a supplementary metric was extracted from the flow field: the fraction of the outlet channel width carrying forward flow in the local channel-parallel direction. As the secondary separation bubble grows, it occupies a larger portion of the channel cross-section, reducing this fraction; when it collapses, the fraction increases. If the secondary separation bubble is indeed responsible for the volume flux variation, the phase of this forward flow fraction phasor at the fundamental frequency should match that of the volume flux phasor. This comparison is shown in Fig.\,\ref{fig:phase of phasors of qFlux and fraction of forward flow} for the $w_\text{o} = 1.4$\,mm case. The agreement is good for distances up to approximately 9\,mm from the splitter, beyond which the forward flow fraction diverges from the volume flux phase, most prominently in the lower channel. This divergence is attributed to an elevated uncertainty in the phase curve in Fig.\,\ref{fig:phase of phasors of qFlux and fraction of forward flow} for $s \geq 9$\,mm resulting from the small magnitudes of the volume flux phasor at these locations (see $w_\text{o}=1.4$\,mm in Fig.\,\ref{fig:q_RMS vs several wo values}). The small magnitudes themselves are explained by Fig.\,\ref{fig:Outlet channel secondary separation bubble at 1.4mm}, which shows how the far end of the secondary separation bubble only reaches $s \geq 9$\,mm when it is at its most expanded state between $\phi = 36^\text{o}$ and $84^\text{o}$ and where it is only a small proportion of the channel width.

The similarity of the corresponding phase-space plots of the upper-minus-lower difference in volume flux per unit depth and forward flow fraction (Figs.\,\ref{fig:Phase-space plot of difference in volume flux} and \ref{fig:Phase-space plot of difference in fractional width of forward flow}) provides further support. In each plot the colour represents the magnitude of the difference; a convecting structure appears as a diagonal band of alternating sign, reflecting the increasing phase lag with streamwise distance, which is precisely what is observed in both plots. The agreement between the two phase-space plots, combined with the phasor comparison in Fig.\,\ref{fig:phase of phasors of qFlux and fraction of forward flow}, supports the interpretation that the growth and decay of the secondary separation bubble is the convecting structure responsible for the streamwise volume flux variation observed for $w_\text{o} \leq 1.4$\,mm. Although a secondary separation is also visible for $w_\text{o} > 1.4$\,mm (see, for example, Fig.\,\ref{fig:exp34 to 41 2D PIV inside oscillator}(j)), its effect on the differential outlet volume flux is minor in those cases, as evidenced by the nearly flat phase in Fig.\,\ref{fig:magnitude and phase of differential volume flux}(b), indicating that the spatially-averaged component dominates. 

\begin{figure}[h]
    \centering    
    \includegraphics[width=0.6\textwidth]{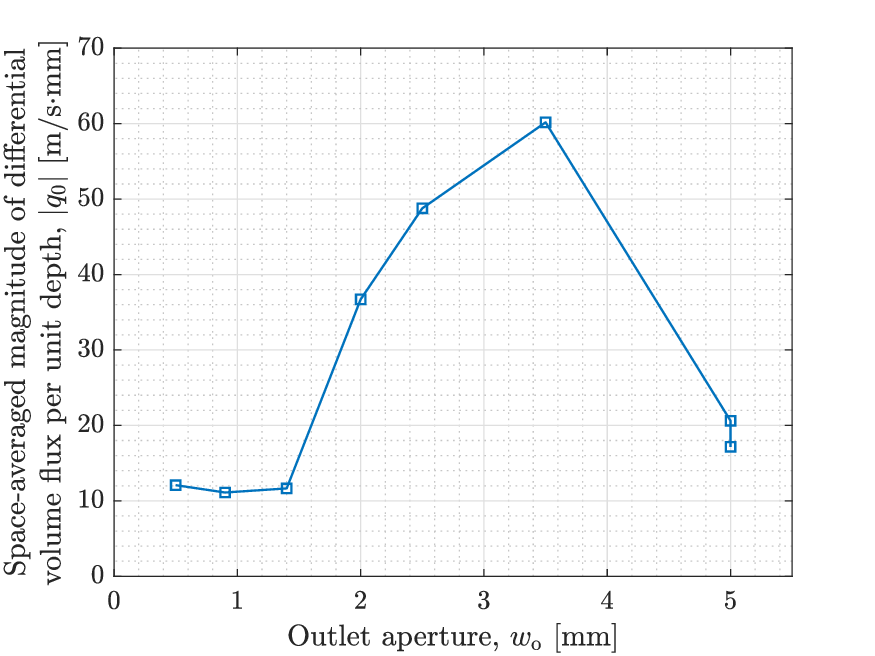}
    \caption{Spatially-averaged differential volume flux per unit depth vs outlet aperture. The two points at $w_\text{o}=5.0$\,mm correspond to the cases where the jet is attached to the upper and lower walls, respectively.}
    \label{fig:Space-averaged volume flux per unit depth vs outlet aperture}
\end{figure}

The spatially-averaged component $q_0$ is the quantity of interest, since it is this component that generates a net differential back pressure. To extract it, a simple convective model was fitted to $q_{\Delta}(s)$ for $w_\text{o} \leq 1.4$\,mm:
\begin{equation}
\label{eq:convection qFlux model}
q_{\Delta}(s) = q_0 + q_\text{c} e^{ik_ss}
\end{equation}
where $q_0, q_\text{c} \in \mathbb{C}$ are the spatially-averaged and convecting components respectively, and the spatial wavenumber $k_s$ was determined from the gradient of a linear fit to the phase vs distance plots in Fig.\,\ref{fig:magnitude and phase of differential volume flux}(b). Stacking the scalar equation at each streamwise location gives the overdetermined linear system
\begin{equation}
\label{eq:convection qFlux model Ax=b form}
\mathbf{q} = \begin{bmatrix} 1 & e^{ik_s s_1} \\ 
                              1 & e^{ik_s s_2} \\ 
                              \vdots & \vdots \end{bmatrix} 
             \begin{bmatrix} q_0 \\ q_\text{c} \end{bmatrix}
\end{equation}
where $\mathbf{q}$ contains the phasor $q_{\Delta}(s)$ measured at each streamwise location. This system was solved in a least-squares sense to extract $q_0$. For $w_\text{o} > 1.4$\,mm, where no convective component was evident, $q_0$ was estimated by ensemble averaging $\mathbf{q}$ directly.\footnote{At the smallest apertures the differential flux magnitude is small and the phase becomes sensitive to the convective correction; accordingly only the magnitude $|q_0|$ is considered here.} The magnitude $|q_0|$ is shown in Fig.\,\ref{fig:Space-averaged volume flux per unit depth vs outlet aperture}.
Fig.\,\ref{fig:Space-averaged volume flux per unit depth vs outlet aperture} reveals a non-monotonic dependence of $|q_0|$ on $w_\text{o}$. At $w_\text{o} = 5$\,mm, $|q_0|$ is small because the jet remains attached to one wall, leaving little oscillatory differential flux. As $w_\text{o}$ decreases to $3.5$\,mm, the back pressure difference is sufficient to initiate the oscillation and $|q_0|$ rises sharply, reaching a maximum at $w_\text{o} = 3.5$\,mm. Further reduction of $w_\text{o}$ causes $|q_0|$ to decrease, becoming small for $w_\text{o} \leq 1.4$\,mm. This non-monotonic behaviour is explained by the onset of the secondary separation: at apertures below $3.5$\,mm the secondary separation increasingly limits the differential outlet mass flux, as described in Section\,\ref{sec:Physical mechanism}, decoupling the outlet flow split from the upstream jet attachment. In spite of the small $|q_0|$ at small apertures, the oscillation remains strong --- the total signal standard deviation $\sqrt{\Sigma\sigma^2}$ is large (Fig.\,\ref{fig:POD mode singular values for several outlet restrictions}) --- confirming that the link between oscillation strength and differential outlet volume flux has been broken.
\subsubsection{Physical mechanism}
\label{sec:Physical mechanism}

The secondary separation bubble plays an additional and distinct role beyond 
introducing a streamwise variation in the differential outlet volume flux: it 
modifies the back pressure mechanism in a way that prevents elimination of the 
oscillation at small $w_\text{o}$. The mechanism is illustrated in 
Fig.\,\ref{fig:annotated flow snapshot wo = 0.5mm}, which shows a flow snapshot 
from the $w_\text{o}=0.5$\,mm case where the flow is attached to the upper wall 
and the secondary separation bubble is large.

When the jet is attached to the upper wall, the greater mass flux in the upper outlet channel produces a higher dynamic pressure recovery at the restriction, generating a back pressure that acts across the jet at the splitter in the direction that opposes the upstream Coand\u{a} attachment --- that is, pushing the jet away from the upper wall. This back pressure causes the jet to separate from the upper wall downstream, near the sharp change in wall curvature upstream of the splitter. The separated flow then bends downward and attaches to the splitter tip, with the secondary separation bubble occupying the region between the separated flow and the upper wall. Two consequences of this behaviour prevent the back pressure from suppressing the oscillation:
\begin{figure}[h]
    \centering    
    \includegraphics[width=0.65\textwidth]{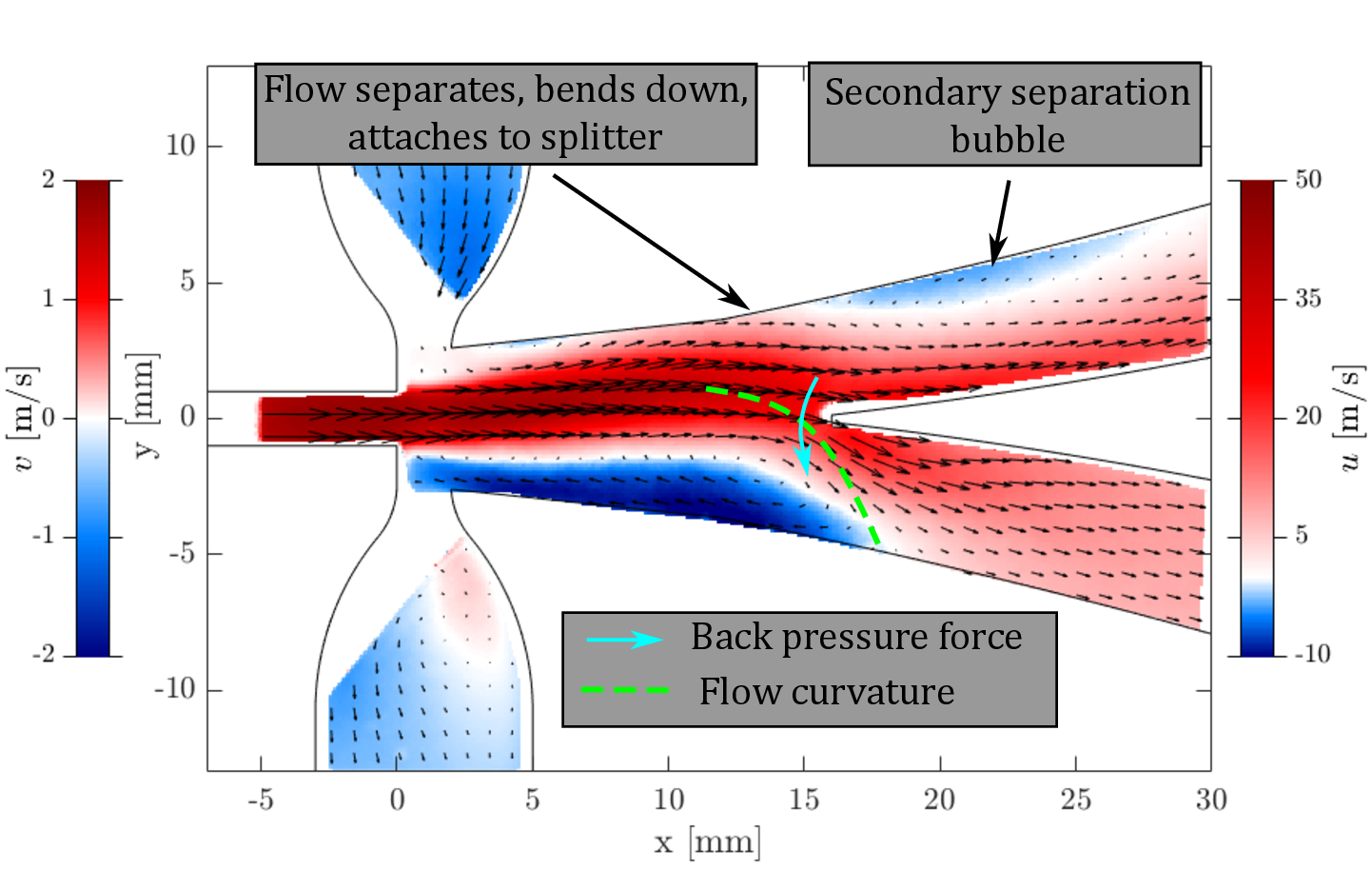}
    \caption{Flow snapshot at $\phi = 24^\text{o}$, $w_\text{o}=0.5$\,mm, 
    showing the secondary separation bubble, the attachment of the separated 
    flow to the splitter, the direction of the differential back pressure 
    force acting on the jet and the curvature of the bending streamlines. Note that flow restriction is further downstream and out of view.}
    \label{fig:annotated flow snapshot wo = 0.5mm}
\end{figure}
\begin{enumerate}

\item \textit{Self-limitation of the back pressure.} Once the deflected jet impinges on the splitter, the flow split between the upper and lower outlets becomes more balanced. This 
directly reduces the differential outlet mass flux, which reduces the 
differential dynamic pressure recovery at the restriction, and hence reduces 
the back pressure difference itself. The back pressure is therefore 
self-limiting: the stronger it grows, the more it redirects flow to the 
splitter, and the more it erodes its own driving source. This prevents the back pressure from growing large enough to overcome the primary Coand\u{a} attachment upstream.

\item \textit{Shielding curvature.} As the streamlines bend downward towards the splitter, they follow a curved path. By the same centripetal pressure balance that governs the Coand\u{a} effect itself, $\Delta p = J/R$, the pressure decreases toward the centre of curvature along the curved streamlines.
The pressure gradient required to maintain the curved streamlines accounts for the pressure difference that would otherwise propagate upstream. In other words, the back pressure is `used up' locally in sustaining the curvature rather than being available to act upstream on the primary attachment, which is thereby shielded from its full influence.

\end{enumerate}
Together, these two effects ensure that as $w_\text{o} \to 0$ the oscillation persists: the self-limiting mechanism prevents the back pressure from becoming arbitrarily large, while the shielding curvature prevents whatever back pressure does exist from reaching the primary attachment region.

\section{Conclusions}
\label{sec:Conclusions}
This paper has investigated two factors that influence the oscillation mechanism of a sonic fluidic oscillator: the geometry of the feedback channel connections at the control ports and the effect of restrictive outlet flow paths. Phase-averaged planar PIV measurements of the internal flow field, analysed using proper orthogonal decomposition, provide the primary evidence. The principal finding is that the assumed direct link between upstream jet attachment and outlet flow split is broken in the present geometry, both by the tendency of the jet to attach to the splitter and by the secondary separation induced by outlet restriction --- a result that has significant implications for the design and application of fluidic oscillators with downstream flow impedances.

The POD analysis demonstrates that the oscillator dynamics are governed by two primary coupled modes: a Sweeping Mode that describes the jet's lateral displacement, and a Bending Mode that describes its curvature during detachment and reattachment.
These modes form a minimal dynamical system: the Sweeping Mode describes the jet's lateral position, while the Bending Mode, lagging by $90^\text{o}$, drives the jet across the device during the switching process. This phase lag arises from the feedback channel dynamics and ensures that there is sufficient feedback flow to sustain the transverse motion of the jet when it is crossing the centreline.
Importantly, modulation of the differential outlet mass flux is governed primarily by the Bending Mode rather than by the lateral jet displacement. This is a consequence of the jet's tendency to attach to the splitter in the present geometry, which weakens the direct relationship between the upstream jet position and the outlet flow split.

A flow structure at the feedback channel entrances plays a critical role in determining the oscillation strength. A time-mean recirculating flow develops at the entrances due to separation induced by the sharp turning of the flow entering the perpendicular feedback channels. This recirculation reduces the effective flow area, limiting the mass flux available to aspirate the separation bubbles and drive the oscillation. The local Reynolds number at the channel entrances lies in a sensitive regime, so that increasing the inlet mass flow rate strengthens the recirculation and progressively throttles the feedback flow, reducing the oscillation strength. Although amplified in the present device by the particular geometry of the channel connection, this throttling effect is expected to persist in any perpendicular feedback channel configuration since the flow must still undergo a sharp turn at entry.

The effect of outlet restriction reveals behaviour that departs significantly from expectations. Rather than suppressing jet attachment as the aperture is reduced, strong oscillations persist down to the smallest apertures investigated ($w_\text{o} = 0.5$\,mm, a contraction ratio of 14.4). At small apertures, a secondary separation bubble forms in the active outlet channel, driven by the differential back pressure acting on the jet near the sharp change in wall curvature upstream of the splitter. This bubble both self-limits the back pressure by equalising the outlet flow split, and introduces a shielding curvature that locally absorbs the back pressure gradient and prevents its upstream propagation to the primary attachment region.

The results suggest several directions for future work and design guidance. The throttling effect of the perpendicular control port connection could be eliminated by orienting the feedback channels parallel to the inlet nozzle, as in the geometry studied by \citet{mcgeachy1973astudy,mcgeachy1973bstudy}. This is consistent with the steady aspiration-switching mechanism \citep{muller1964study}, which depends on the mass flux through the feedback channel rather than on the momentum or direction of the incoming flow. The decoupling between upstream jet attachment and outlet flow conditions identified here, while undesirable in an oscillator context, may be beneficial in fluidic diverter applications where robustness to large transient differential back pressures --- such as those induced by passing turbomachinery or flow structures at the outlet --- is required.
\section*{Acknowledgements}
This work was supported by ONR Grant N00014-21-1-2293, monitored by Dr.\,Leighton Myers and Dr.\,David Gonzalez, and the ATI grant Liquid Hydrogen Gas Turbine (LH2GT), UKRI project reference 113263.

\bibliography{references_fixed.bib}

\end{document}